\documentclass[11pt,tightenlines,eqsecnum,floats,aps,amsmath,amssymb,nofootinbib,prd,shownopacs]{revtex4}

\usepackage{amsmath,amssymb,amsfonts,amsthm}
\usepackage{graphicx}
\usepackage{enumerate} 
\usepackage{colordvi} 

\usepackage[mathscr]{eucal}
\usepackage[latin1]{inputenc}
\usepackage{amsmath}
\usepackage{amsfonts}
\usepackage{amssymb}
\usepackage{graphicx}
\usepackage{enumerate}
\def\lp{{\ell}_{\rm Pl}}

\def\q{\mathring{q}}
\def\e{\mathring{e}}
\def\w{\mathring{\omega}}

\newcommand{\f}{\frac}

\newcommand{\fref}[1]{Fig.\,\ref{#1}}

\DeclareMathOperator*{\sgn}{sgn}

\def\f{\frac}

\usepackage{enumerate}

\newcommand{\be}{\nopagebreak[3]\begin{equation}}
\newcommand{\ee}{\end{equation}}
\newcommand{\ba}{\nopagebreak[3]\begin{eqnarray}}
\newcommand{\ea}{\end{eqnarray}}

\newcommand{\bmult}{\nopagebreak[3]\begin{multline}}
\newcommand{\emult}{\end{multline}}

\def\lp{{\ell}_{\rm Pl}}

\def\sinamu{\frac{\sin{(\bar{\mu}_1 c_1)}}{\bar{\mu}_1}}
\def\sinb{\sin{(\bar{\mu}_2 c_2)}}
\def\sinbmu{\frac{\sin{(\bar{\mu}_2 c_2)}}{\bar{\mu}_2}}
\def\sinc{\sin{(\bar{\mu}_3 c_3)}}

\def\cosa{\cos{(\bar{\mu}_1 c_1)}}

\def\Heff{\mathcal{H}_{\rm eff}}
\def\Hmatt{\mathcal{H}_{\rm matt}}
\def\mua{\bar{\mu}_1}
\def\mub{\bar{\mu}_2}
\def\muc{\bar{\mu}_3}
\def\lp{l_{\rm Pl}}

\begin{document}

\title{Quantum gravitational Kasner transitions in Bianchi-I spacetime}
\author{Brajesh Gupt}
\email{bgupt1@lsu.edu}

\author{Parampreet Singh}
\email{psingh@phys.lsu.edu}
\affiliation{Department of Physics and Astronomy, Louisiana State University, Baton Rouge, 70803}

\pacs{04.60.Pp, 04.60.Kz, 98.80.Qc}

\begin{abstract}
Due to non-perturbative quantum gravitational effects, the classical big bang singularity is replaced by a quantum big bounce of the mean scale factor in loop quantization of Bianchi-I spacetime.  An important issue is to understand various differences in the physical properties of the spacetime across the bounce. We investigate this issue in the context of various geometrical structures, identified by the Kasner exponents of the metric, which arise on approach to the singularity
in the classical theory. Using effective spacetime description of Bianchi-I model in loop quantum cosmology with dust, radiation and stiff matter, we find that as in the classical theory, geometrical structures such as a cigar or a pancake form, but they are finite and non-singular. Depending on the initial conditions of the matter and anisotropies, different geometric structures are possible in the pre- and post-bounce phases in physical evolution. Thus, quantum gravitational effects can cause a Kasner transition in Bianchi-I spacetime, which is not possible at the classical level. Interestingly, we find that not all transitions are allowed at the level of effective dynamics in loop quantum cosmology. We find the selection rules and underlying conditions for all allowed and forbidden transitions. Selection rules suggest that for a given set of initial conditions on anisotropies, occurrence of Kasner transitions follows a distinct pattern, and certain transitions are  more favored than others.
\end{abstract}

\maketitle


\section{Introduction}
Singularities such as the big bang are the boundaries of spacetime in Einstein's theory of  general relativity (GR). Signalling the limits of validity of GR, and generally characterized by  inextendibility of geodesics and divergence of curvature scalars, they come in a diverse variety. They  can be  strong or weak (harmless) \cite{tipler77,krolak86}, and  can arise  by divergence of Ricci or Weyl curvature, or both. They may have various   geometries such as a barrel, cigar, pancake or a point, as in a Bianchi-I spacetime \cite{dorosh,thorne1,jacobs2,ellismac}. These geometrical structures are  described via Kasner exponents in the spacetime metric, and it has been conjectured by Belinski-Khalatnikov-Lifshitz (BKL), that 
a general approach to singularity may involve chaotic oscillations between different Kasner phases \cite{bkl, misner2}. However, for the matter satisfying weak energy condition, such transitions require a spacetime to have both spatial curvature and anisotropies,  hence, these are absent in a classical 
Bianchi-I spacetime.  For such spacetimes,  the geometric structure of the  singularity is  determined from the initial conditions on the matter content and anisotropies, through the resulting interplay between the  Ricci and Weyl parts of the spacetime curvature. As an example, for a Bianchi-I universe in GR, filled with dust, only cigar and pancake structures are possible, whereas if filled with stiff matter, barrel, cigar and point like singularities can form \cite{jacobs2}.

Such singularities are expected to be resolved in a quantum theory of gravity. Though the exact mechanism of this resolution is still to be understood for arbitrary spacetimes in quantum gravity, analysis of various isotropic and anisotropic models based on loop quantum gravity (LQG) suggests that the singularities are replaced by a non-singular bounce which occurs due to  underlying discrete quantum geometry, and bridges two classically disjoint branches of  continuum spacetime. The occurrence of bounce, first obtained for spatially flat isotropic models in loop quantum cosmology (LQC) for massless scalar field (which is equivalent to stiff matter in terms of equation of state) \cite{aps1,aps2,aps3},  has been established in various matter models \cite{apsv,bp,acs,kv,szulc,kp1,ap} and in the presence of anisotropies \cite{awe2,awe3,we1,madrid1,madridb1,madrid_ed}.\footnote{For some of the earlier attempts to quantizing the anisotropic spacetimes based on the old quantization methods see eg. \cite{bdh,bdprl,chioub1}. See also Ref. \cite{bck07} for a discussion on von-Neumann stability issues in anisotropic models in LQC.} For an up to date review of recent results, see Ref. \cite{as1}. A robust feature of these models is that non-perturbative quantum gravitational effects lead to singularity resolution when spacetime curvature becomes Planckian, irrespective of the initial conditions. These effects occur in a very short range of spacetime curvature, and quickly die when spacetime curvature becomes small, roughly at a percent of the Planck value below which LQC approximates GR.   Various physical insights on quantum geometry have been obtained using  effective continuum spacetime description of LQC  \cite{jw,vt,psvt}, which provides an accurate representation of the underlying discrete quantum dynamics for states which describe a macroscopic universe at late times \cite{aps3,bp,ap,szulc_b1,madridb1,dgs}. Using effective dynamics, in context of the details of singularity resolution, bounds on the energy density, expansion and shear scalars in different models have been found \cite{gs1}, and strong singularities have been shown to be generically resolved in isotropic \cite{ps09,sv} and Bianchi-I models \cite{ps11}. For more discussion on semiclassical limit and effective dynamics see Sec.-V of Ref. \cite{as1} and the references therein.

These developments set a stage of posing various interesting questions which fall beyond the realm of the classical theory. Of these perhaps one of the most important one concerns with understanding the way physics in the pre-bounce branch  influences the physics in the post-bounce branch (and vice-versa).  For the isotropic models in LQC, useful insights have been gained by considering properties of semi-classical states when evolved across the bounce \cite{cs08,kp2,cm1}. 
As the bounce is approached, the geometrical structure of the spatial geometry is always point like on both sides of the bounce in the isotropic and homogeneous models.
These results can change in anisotropic setting where due to the presence of Weyl curvature, physics is much richer. 
In particular, given that in Bianchi-I spacetime, the classical theory leads to different geometric structures as the singularity is approached, it becomes important to understand how such structures arise in effective spacetime description of LQC and the way they are affected across the bounce.
 It is important to note that far away from the Planck regime, since there is little distinction between LQC and GR, for a given set of initial conditions on the matter and 
anisotropies, a co-moving observer in GR and another  in LQC measure practically the same Kasner exponents of the spacetime metric. Such observers thus deduce the formation of the same geometric structure, such as a cigar or a pancake, in high curvature regime.  The important difference between their description is, that the observer in GR, is inevitably annihilated as the geometric structure becomes singular at the big bang, whereas the observer in LQC finds geometric structure to be finite and singularity free.

These and related aspects of new physics have so far remained unexplored in the studies on Bianchi models in LQC.\footnote{For some  of the phenomenological implications see  for eg. \cite{cv,csv,cs09,roy_bianchi1,artym_bianchi1,cm2}.}
Our goal in this work is to investigate these issues for Bianchi-I spacetimes where detailed results on the formation  of geometric structures at the classical level are already available \cite{jacobs2}. We consider effective dynamics of Bianchi-I model in LQC for a barotropic fluid
 \footnote{In the loop quantization of perfect fluid there are several open issues such as that of deparametrization. See eg. Ref. \cite{dgkl} for a discussion. In this analysis, our treatment of perfect fluid will be at an effective Hamiltonian level.
 } 
 with equation of state corresponding to stiff matter, dust and radiation, with an objective to understand answers to the following questions:
Given a set of initial conditions under what conditions do particular geometric structures arise in LQC as one approaches the bounce of the mean scale factor? Does a physical evolution across the bounce yield the same geometric structure in the pre-bounce and post-bounce phases? If not, then is such a  transition across the bounce random, or does it follow a set of rules? If such rules exist, under what conditions are some transitions favored over others, and finally how do they depend on the matter content?

In our  presented analysis, based on effective spacetime description, we outline all the conditions under which particular geometric structures form in the Bianchi-I model in LQC. This result stands on an analog footing as the work of Jacobs' in the classical theory \cite{jacobs2}. We show that there exists a hierarchy amongst different structures. Whether or not a certain geometric structure forms, depends on the ratio of shear and the expansion scalar. For large values of expansion  normalized shear, cigar structures are more probable, whereas for small values of ratio of shear and expansion scalar, a point like or isotropic  approach to the bounce occurs.
Moreover, depending on the initial conditions, pre-bounce and post-bounce branches may have different geometric structures. That is, a universe which forms a cigar type structure in the contracting branch before the bounce, may form cigar, barrel, pancake or point like structure in the expanding branch (in the backward evolution). Interestingly, such a Kasner transition is not random. It follows selection rules which are determined by the initial relative 
strength of the matter and anisotropies, captured in terms of two anisotropy parameters. The  Kasner transitions which take place across the bounce in Bianchi-I model in LQC remind us of the mixmaster behavior between Kasner phases in classical Bianchi-IX universe. However, there are important differences. As mentioned earlier, there are no Kasner transitions in the Bianchi-I model in GR. Thus, such transitions are purely quantum gravitational in origin. Further, unlike the Kasner transitions in the classical Bianchi-IX spacetime, there is only one transition across the bounce. And finally, the Kasner transitions in the classical Bianchi-IX model lie on the same branch (expanding or contracting) of the universe, whereas in LQC, they involve a non-singular transition between expanding and contracting branches. 

The plan of this work is as follows. In the following section, we start with the classical GR to provide a stage for discussion of the new results in LQC. 
We discuss the derivation of the primary dynamical equations in classical Bianchi-I spacetime using classical Hamiltonian constraint, and illustrate various geometric structures using the Kasner exponents. Here we provide a hierarchy among different geometric structures, which to our knowledge has not been discussed before in the classical theory. We discuss parameterization introduced by Jacobs \cite{jacobs2} which proves very useful in the analysis of the formation of geometric structures.  This is followed by a summary of the results in the case of stiff matter or the Zel'dovich solution, and dust and radiation. In Sec. III,  we start with the effective Hamiltonian constraint of Bianchi-I model in LQC,  and summarize the derivation of the effective dynamical equations which are then numerically solved. We analyze the cases of stiff matter, and dust and radiation separately, and classify all kinds of geometric structures and the allowed and forbidden transitions across the bounce. Defining a ratio in the parameter space, we also  compute how favored a particular type of transition is for a given equation of state and the anisotropic shear scalar. We conclude with a summary of main results in Sec. IV.

\section{Classical theory: Dynamics, Anisotropic Hierarchy, Jacobs' parameterization and Examples}

Since loop quantization is based on classical gravitational phase space variables, the Ashtekar-Barbero connection $A_a^i$ and the triads $E_i^a$, in this 
section we start with a discussion on the classical theory in connection variables, and then relate it to conventional metric variables.  In the first part of this section 
we obtain the dynamical equations resulting from the classical Hamiltonian constraint and discuss the Kasner exponents obtained from integration of the 
dynamical equations. We then use these Kasner exponents to elucidate the structure of the singularities in classical Bianchi-I model and establish a hierarchy amongst them using expansion normalized shear scalar. A useful way to understand the properties of the anisotropic structure of the spatial geometry of spacetime is via parameters introduced by Jacobs \cite{jacobs2}. After discussing these parameters, which we later use to classify the Kasner transitions, we illustrate some key properties of Bianchi-I spacetime with stiff matter (or massless scalar), and dust and radiation.

\subsection{Dynamical equations}
We consider a homogeneous and diagonal Bianchi-I spacetime with a manifold $\Sigma\times\mathbb R$ where $\Sigma$ refers to the flat spatial manifold having a ${\mathbb R}^3$ topology. Due to the non-compact nature of the spatial manifold, one needs to introduce a fiducial cell  $\mathcal V$ in order to define the symplectic structure. The fiducial cell is chosen to have a fiducial volume $V_o=l_1l_2l_3$ where $l_i$ are the coordinate lengths in the three spatial directions. The edges of the cell are chosen to lie along fiducial triads $\e_a^i$.  We denote the fiducial metric compatible with the co-triads $\w^i_a$ as  $\q_{ab}$. Using the underlying symmetries of the homogeneous spatial manifold, the Ashtekar-Barbero connection $A_a^i$ and the triads $E_i^a$, can be symmetry reduced as:

\be
A_a^i \, = \, c^i \, (l_i)^{-1} \, \w_a^i, ~~ \mathrm{and} ~~ E_i^a \, = \, p_i \, l_i\, V_o^{-1}\, \sqrt{\q} ~,
\ee
where $i = 1,2,3$. The symmetry reduced connection $c^i$ and triad $p_j$ satisfy the following Poisson bracket relation
\be
 \{c^i,p_j\}=8\pi G\gamma \delta^i_{j}
\ee
where $\gamma=0.2375$ is the Barbero-Immirizi parameter fixed by the black-hole entropy calculation in LQG.  The triads are related to the directional scale factors $a_i$ in the  Bianchi-I metric
    \be
\label{metric}  d s^2\, =\, -d t^2 + a_1^2\, d x^2 +a_2^2\, d y^2 +a_3^2\, d z^2
  \ee
as 

\be
p_1 \, =\,  \, l_2\, l_3\, |a_2 \, a_3|, ~~ p_2\, =\,  l_1 \, l_3 \, |a_1 \, a_3|, ~~ p_3 \, = \,  l_2 l_1 |a_1 a_2|.
\ee

The classical Hamiltonian constraint for Bianchi-I spacetime in terms of the above connection and triad variables, for lapse $N = 1$, is  \cite{chioub1}
\be
\mathcal{H}_{\rm{cl}}=\frac{1}{8 \pi G \gamma^2\, V} \, \left(c_1 p_1 c_2 p_2 + \mbox{cyclic terms}\right) +\Hmatt ~,
\ee
where without any loss of generality we have chosen fiducial lengths $l_i$ to be unity, $V=\sqrt{p_1p_2p_3}$ is the physical volume of the cell ${\cal V}$ and $\Hmatt$ denotes the matter part of the Hamiltonian. 
We consider the matter as a perfect fluid with a barotropic equation of state $w = P/\rho$, where $P$ and $\rho$ denote the pressure and the energy density of the 
matter content.

The dynamical equations for the triad and the connection components can be derived using the Hamilton's equations:
\be
{\dot p}_i \, = \, \{p_i,{\cal H}_{\rm cl}\} \, = \, -8 \pi G \gamma \f{\partial \cal{H}_{\rm cl}}{\partial c_i}, ~~~~ {\dot c}_i \, =\, \{c_i,{\cal H}_{\rm cl}\} \, = \, 8 \pi G \gamma \f{\partial \cal{H}_{\rm cl}}{\partial p_i} 
 \ee
  where a `dot' refers to the derivative with respect to the proper time $t$. Using these equations, it is straight forward to obtain the classical evolution equations 
for Bianchi-I spacetime
\be\label{dyneq1}
H_1H_2+H_2H_3+H_3H_1=8\pi G \rho
\ee
and 
\be
\label{dyneq2}{\dot H}_2+{\dot H}_3+H_1^2+H_2^2+H_3^2=-8\pi G P
\ee
(and its cyclic permutations).  Here $H_i$ denote the directional Hubble rates, which are related to the time derivatives 
of the triad components, such as 
  \be
\label{h1def} H_1\, = \, \f{\dot a_1}{a_1} = \, \f{1}{2} \left(\f{{\dot p}_2}{p_2} + \f{{\dot p}_3}{p_3} - \f{{\dot p}_1}{p_1}\right), 
 \ee
(and similarly for  $H_2\, {\rm and}\, H_3$). The energy density and pressure in the above equations, satisfy the conservation law

\be
\dot{\rho} = -3H(\rho+P),
\ee
where $H=(H_1+H_2+H_3)/3$ is the mean Hubble rate. Using the relation $P = w \rho$, the energy density can be expressed as follows
\be
\rho(t)=\rho_o\, a(t)^{-3(1+w)}~,
\ee
where $\rho_o$ is the initial energy density, $a=(a_1a_2a_3)^{1/3}$ is the mean scale factor and $a(t=0) = 1$.

We now consider the parts of the expansion matrix in the Bianchi-I spacetime. For the diagonal Bianchi-I metric, the expansion tensor $\theta_{\alpha \beta}$ 
is given by the covariant derivative of the unit fluid velocity, with its trace defining the expansion scalar $\theta$ of a co-moving observer:
\be
\label{thetadef}\theta = H_1+H_2+H_3 ~.
\ee
The traceless part of the expansion matrix defines the anisotropic shear which takes the following form for the diagonal Bianchi-I spacetime:
\be
\sigma_{ij} = {\rm diag}\left( \sigma_1,\, \sigma_2,\, \sigma_3\right)
\ee
where $\sigma_i = H_i-\f{1}{3}\theta$ are the anisotropy parameters. Since the shear matrix $\sigma_{ij}$ is traceless, by definition its components satisfy the  following constraint
\be
\label{traceless}\sigma_1+\sigma_2+\sigma_3=0.
\ee

An important measure of the anisotropy of the spacetime is the shear scalar, defined as $\sigma^2=\sigma^{ij}\sigma_{ij}$, which can be expressed in terms 
of $H_i$ as
\be
\label{sheardef}\sigma^2= \sum_i \sigma_i^2 = \f{1}{3}\left((H_1-H_2)^2+(H_2-H_3)^2+(H_3-H_1)^2\right).
\ee

In the classical theory, using the dynamical equations it follows that $\sigma^2 \propto a^{-6}$. Further, it satisfies the generalized Friedmann equation:
\be
H^2 = \f{8 \pi G}{3} \rho ~+ ~ \f{1}{6}\sigma^2 ~.
\ee

In the absence of the matter, solution of the dynamical equations (\ref{dyneq1}) and (\ref{dyneq2}), gives the Kasner solution according to which the directional scale factors can be written as
\be
\label{a1kasner}a_i \propto t^{k_i}
\ee
where constants $k_i$ are the Kasner exponents. For vacuum Bianchi-I, these exponents satisfy the following two constraints
\be
\label{kasner1}k_1+k_2+k_3= 1 ~~ \mathrm{and} ~~ k_1^2+k_2^2+k_3^2=1.
\ee
\vskip0.3cm
For general matter, the directional scale factors can not be expressed in terms of eq.\ (\ref{a1kasner}) with (\ref{kasner1}) holding true. However, for the case of $w=1$, it is possible to write similar constraints, given by \cite{bk73}
 \be
\label{kasner2}k_1+k_2+k_3= 1 ~~ \mathrm{and} ~~ k_1^2+k_2^2+k_3^2=1 - k^2,
\ee
where $k^2$ is a constant which can be expressed in terms of the energy density and shear scalar as $k^2 = \f{2}{3}\left(\f{8\pi G \rho/3}{8\pi G \rho/3+\sigma^2/6}\right)$. For a massless scalar field (for which the equation of state is $w=1$) the constant $k^2$ can also be expressed in terms of its field momentum ($P_\phi$, which is a constant of motion) as-

\be
k^2=  \f{2}{3}\left(\f{4\pi G P_\phi^2/3}{4\pi G P_\phi^2/3+\Sigma^2}\right)
\ee

where, $\Sigma^2=\f{1}{6}\sigma^2 a^6$ is also a constant motion.

We now have the expression for $k^2$ in terms of the two constants of motion of the system for a massless scalar field. It is evident from the above equations that its value depends on the initial data and $k^2 \leq 2/3$. The Kasner exponents in terms of the directional Hubble rates can be given as $k_i= H_i/(3H)$. Utilizing this relation and $\sigma_i = H_i-H$, it is straightforward to show that the constraints on the Kasner exponents given in eq.\ (\ref{kasner2}) follow from eqs.\ (\ref{traceless} and \ref{sheardef}). The second constraint in eq.\ (\ref{kasner2}) can also be derived from the classical Hamiltonian constraint.

 The case of massless scalar field is interesting as it is the only matter, 
satisfying dominant energy condition, for which Bianchi-I universe does not isotropize as the universe expands. This is so because the energy density for $w=1$ scales as 
$\rho \propto a^{-6}$, in the same way as $\sigma^2$. Behavior of the energy density in relation to $\sigma^2$ also shows that unless the matter is massless scalar field, the anisotropic shear always dominates $\rho$ as the classical singularity is approached (for $w \leq 1$). Thus, for $w < 1$, a Bianchi-I spacetime near the singularity approaches the Kasner vacuum solution.

\subsection{Structure of the singularities and their hierarchy}

Presence of the anisotropic shear makes the structure of the spacetime near the singularity much richer in comparison to the 
isotropic spacetime. Recall that in an isotropic universe, the big bang singularity is always point like. However, in the Bianchi-I spacetime 
the geometrical nature of singularity depends on whether or not the scale factors in the three directions approach zero together, and can be 
classified as follows \cite{thorne1}

\begin{itemize}

\item {\bf Point:} The approach to singularity is said to be point like when all the scale factors tend towards zero simultaneously i.e. when all three Kasner exponents $k_1,k_2,k_3>0$. These singularities occur when the energy density of the matter content plays an important role, and thus only occur for $w \geq 1$. At these singularities, the anisotropic shear does not play a dominant role. In the above sense, a point singularity is often also termed as isotropic singularity in Bianchi-I model.

\item {\bf Barrel:} Barrel singularity occurs when one of the scale factors tends to a constant while the other two approach zero i.e. $a_1 \rightarrow {\rm const}$ and $a_2, a_3 \rightarrow 0$. The corresponding Kasner exponents for barrel type singularity are $k_1=0$ and $k_2, k_3 > 0$.

\item {\bf Pancake:} Pancake singularity occurs when one of the scale factors tends to zero while the other two approach to constant i.e. $a_1\rightarrow 0$ and $a_2,a_3\rightarrow {\rm const}$ which means that the Kasner exponents $k_1 >0$ and $k_2, k_3 = 0$. This singularity appears as one of the axisymmetric solutions for $0\leq w<1$ and does not take place for $w=1$.

\item {\bf Cigar:} The approach to singularity is classified as cigar when one of the scale factors diverges while the other two approach to zero i.e.\ $a_1\rightarrow \infty$ and $a_2,a_3\rightarrow 0$ as the singularity is approached. For cigar type singularity one of the Kasner exponents is negative while the remaining two are positive i.e. $k_1 < 0$ and $k_2, k_3 \, > \, 0$

\end{itemize}

\begin{figure}[tbh!]
\includegraphics[angle=0,width=0.47\textwidth]{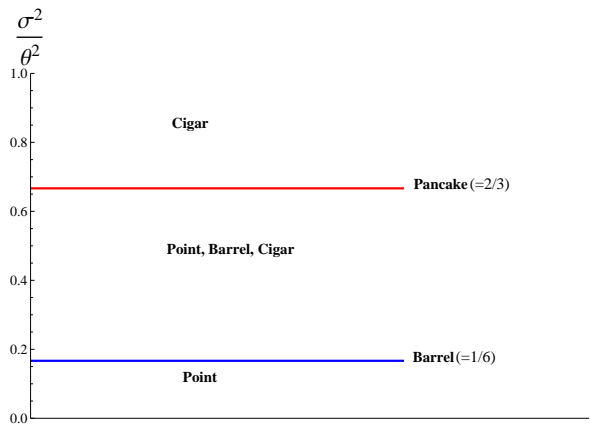}
\hskip0.5cm
\caption{Hierarchy of singularity structure in the classical theory based on $\sigma^2/\theta^2$ for Bianchi-I spacetime.}
\label{hierarchy}
\end{figure}

 It is to be noted that  the approach to singularity for the vacuum Bianchi-I spacetime is always either cigar or pancake like. Introducing the stiff matter (perfect fluid with $w=1$) to the Bianchi-I spacetime brings in other types of the approach to singularity which were otherwise forbidden in the vacuum Kasner solution. The geometric structure of singularity is decided by the anisotropy present in the system. As remarked above, for a point like singularity to exist, the matter part should dominate over the anisotropy. On the other hand if the anisotropy wins over the matter, the point like approach to singularity becomes less likely. 
  
  It turns out that one can define a hierarchy of the various approaches to the singularities based on the value of $\sigma^2/\theta^2$, the normalized dimensionless shear scalar.  
Using the definitions of the shear and the expansion scalar, the normalized shear scalar can be expressed in terms of the Kasner exponents $k_i$ as follows:
\be
\label{Sigma}\f{\sigma^2}{\theta^2} = \f{2}{3}\left(1-3(k_1k_2+k_2k_3+k_3k_1)\right).
\ee
 The ratio $\sigma^2/\theta^2$ takes the smallest value when all three Kasner exponents are positive which is the case for point singularity. It is also evident from the expression for $\sigma^2/\theta^2$ and the definitions of the various singularities that for a pancake singularity $\sigma^2/\theta^2=2/3$, for barrel singularity $1/6 \leq \sigma^2/\theta^2 <2/3$ and for cigar singularity  $\sigma^2/\theta^2>1/6$. 
This hierarchy of the singularities based on the normalized shear scalar is shown in \fref{hierarchy}. From here we can say that the point like singularity is the least  anisotropic structure of singularity while cigar corresponds to highly anisotropic situations.

\subsection{Jacobs' parameterization}

It is convenient to capture the information contained in these, by using parameters introduced by Jacobs Ref. \cite{jacobs2}. One first defines two parameters $\epsilon$ and $\psi$ as:

\be
\label{eps}\epsilon = \sqrt{\f{\sigma^2}{4\pi G \rho}}
\ee
and
\be
\label{psi}  \sin\left(\psi\right)=\sqrt{\f{3}{2}\f{\sigma_1^2}{\sigma^2}}, \quad \sin\left(\psi+\f{2\pi}{3}\right)=\sqrt{\f{3}{2}\f{\sigma_2^2}{\sigma^2}}, \quad \sin\left(\psi+\f{4\pi}{3}\right)=\sqrt{\f{3}{2}\f{\sigma_3^2}{\sigma^2}}.
\ee

We will refer to these transformed parameters as Jacobs' anisotropy parameters in the rest of the article. The ranges of the Jacobs' anisotropy parameters are $|\epsilon|=\left(0,\infty\right)$ and $\psi=\left[0,2\pi/3\right]$. Note that in the classical theory $\epsilon = 0$ is reached only in the isotropic limit of the model.  Similarly, $\epsilon=\infty$ is also reached only for a vacuum Bianchi-I model or when the approach to singularity is dominated by anisotropies.\footnote{In the stiff matter universe, a non-zero finite $|\epsilon|$ in the initial data ensures that $|\epsilon|$ will never be zero or infinity during the evolution, not even if $\sigma^2$ and $\rho$ diverge as the singularity is approached. In LQC, since the energy density and shear both are finite during the entire evolution for Bianchi-I spacetime, $\epsilon=0$ is possible when $\sigma^2=0$, which is the isotropic limit of the model and $\epsilon=\infty$ takes place only when $\rho=0$ which is only allowed for the vacuum case.}

 It is evident from the eqs.\ (\ref{eps} and \ref{psi}) that $\epsilon$ refers to the ratio of  the total anisotropic shear present in the spacetime to the energy density while the parameter $\psi$ tells how the total anisotropy is distributed in the three directions. Thus, one has complete information about the shear matrix $\sigma_{ij}$ of the spacetime, when both $\epsilon$ and $\psi$ are specified.

The classical dynamical equations in GR can be analytically integrated for constant equations of state $w$. For $0\leq\,w\,\leq1$, upon integration, the Einstein's equations give the following expressions for the scale factors in terms of Jacobs' parameters \cite{jacobs2}
\be
\label{scalefact} a_i(t) \propto \begin{cases}
                     a(t)\left( \f{\left[4\,a(t)^{3(1-w)} + \epsilon^2\right]^{1/2}+|\epsilon|}
                      {\left[4\,a(t)^{3(1-w)} + \epsilon^2\right]^{1/2}-|\epsilon|}\right)^{\f{2\sgn(\epsilon)Z_i}{3(1-w)}} & (0\leq\,w\,<1) \\
                      \\
                      a(t)^{1- 2\,\epsilon Z_i (4+\epsilon^2)^{-1/2}} & (w=1)

                      \end{cases}
\ee

where $$Z_1=\sin\left(\psi\right),Z_2=\sin\left(\psi+\f{2\pi}{3}\right),\,Z_3=\sin\left( \psi+\f{2\pi}{3}\right).$$ Also, the expressions for the directional Hubble rates are given via,
\be
\label{hubble} H_i(t)= \f{\left[4\,a(t)^{3(1-w)} + \epsilon^2\right]^{1/2} \mp 2|\epsilon|Z_i}{3(6\pi \rho_o)^{-1/2} a^3(t)} \quad (0\leq \, w \leq \, 1)        
\ee
where $\rho_o$ is the initial energy density. In the following subsections we discuss the properties of these solutions for the case of massless scalar, dust and radiation.

\subsection{Stiff matter}

The solution of Bianchi-I spacetime with stiff matter is also referred to as Zel'dovich solution\cite{zeldovich62}. For such a universe where $w=1$ the energy density varies with the mean scale factor as $\rho \propto a^{-6}$, i.e. in the same way as the shear scalar $\sigma^2$ in the classical theory.
  Thus, the ratio of the shear scalar to the energy density is a constant, which means that the Bianchi-I spacetime filled with stiff matter never  isotropizes. Further, depending on the initial conditions,  it is also possible that the matter density dominates over anisotropy and the approach to singularity is point like.\ Hence, such a universe does not, in general, behave like vacuum Kasner universe as the singularity is approached. 
For a Bianchi-I universe with stiff matter, $|\epsilon|$ is a constant of motion in the classical theory, and the Kasner like behavior occurs only when the anisotropic contribution to the spacetime curvature dominates over the contribution due to the matter. This requires  $\epsilon$ to be greater than a critical value $\epsilon^*$, where for any $\epsilon < \epsilon^*$, the approach to singularity will be always point like and the Kasner behavior will be absent. We  now find out this critical value $\epsilon$.

We first make the following transformation of the parameter $\epsilon$ \cite{jacobs2}:
\be
\label{deltadef} \delta = -\f{\epsilon}{\sqrt{4+\epsilon^2}}.
\ee
This transformation restricts the range of  new anisotropy parameter $\delta$ to $0 < |\delta| < 1$.  In terms of 
$\delta$ and $\psi$, the expressions for the directional scale factors and Hubble rates  for stiff matter universe take a much simpler form,
\be
a_i(t) = a_{i\,o} \, a(t)^{1\pm 2|\delta| Z_i}, ~~~ \mathrm{and} ~~~ H_i(t) =  \f{(1\pm2 |\delta| Z_i)}{3 a(t)^3} ~.
\ee

Here $a_{i\,o}$ is the directional scale factor at the initial time.
From the above equations, it is clear that for all of the Hubble rates to be positive, which is the requirement for a point like approach to singularity (in the expanding branch), $\delta$ must be given by $1\pm 2|\delta| > 0$.\ Thus,  
for $|\delta| < 1/2$ the approach to singularity will always be point like, irrespective of the value of $\psi$. As we will discuss in the next section,  for $\delta \geq 1/2$, the approach to the classical singularity can be point like, barrel or cigar. Using the definition of $\delta$, we find that the critical value of $\epsilon$ below which the approach to the classical singularity is always point like turns out to be $\epsilon^*= 2/\sqrt 3$. Existence of such a critical value is an  important property of the Bianchi-I spacetime with stiff matter, and  we will see in the next subsection that such a critical value of anisotropy parameter is absent for $w<1$ for which the point like singularity is forbidden.
 Integrating Einstein's equations, the mean scale factor for $w=1$ universe can be written as a function of time as
  \be
a(t) \propto (t+t_s)^{1/3}
 \ee
  where $t_s$ is an integration constant. 
 Now let us write the classical directional scale factors in terms of the Kasner exponents $k_i$ as follows:
  \be
a_i(t) \propto (t+t_s)^{k_i}
 \ee
  where $k_i$ are the Kasner exponents.
The anisotropy parameters $\delta$ and $\psi$ can now be expressed in terms of the directional scale factors and the Kasner exponents as follows -
 
 \be
\label{psidef}\psi (t) \, =\, \tan^{-1}\left(\f{1}{\sqrt{3}}\, +\, \f{2\,\log\left(a_1/a_2\right)}{\sqrt{3}\,\log\left(a_2/a_3\right)}\right)
 \ee and 
 
 \be
\label{deldef}\delta (t)\,=\, \f{3\left(k_1-k_2\right)}{2\left(\sin(Z_1)-\sin(Z_2)\right)}~.
 \ee

\vskip0.5cm
Bianchi-I spacetime with stiff matter admits all kind of singularity except the pancake type, as the Kasner exponents for the pancake singularity are $k_1=1, k_2,k_3=0$ and they do not satisfy the constraint on the Kasner exponent for $w=1$ given in eqs.\ (\ref{kasner2}). The axisymmetric solution could be either cigar or barrel depending on the value of $\delta$ and $\psi$. We notice that if the anisotropy parameter ($\delta$) is small enough, the only possible singularity is point. For higher values of $\delta$, barrel and cigar singularity form depending on the value of the parameter $\psi$. The various possible structure of singularity in the classical Bianchi-I model for stiff matter universe are shown in the second column of Tables \ref{table1} and \ref{table2} in the following section.

\subsection{Dust ($w=0$) and radiation ($w=1/3$)}

Solution for a general equation of state in the range $0\leq w < 1$ is given via eqs.\ (\ref{scalefact} and \ref{hubble}). The energy density depends on the mean scale factor as $\rho \propto a(t)^{-3(1+w)}$ and hence the ratio of the anisotropic shear scalar to the energy density is given via

\be
\label{w01}\f{\rm \sigma^2}{\rm \rho} \propto a(t)^{-3(w-1)}~.
\ee

It is evident from the above equation that the shear always dominates over the energy density for $w<1$ as the singularity is approached i.e.\ $a(t)\rightarrow 0$ which implies that the point like singularity does not occur.  Near the singularity the spacetime behaves similarly to vacuum Bianchi-I given via the vacuum Kasner solution. Moreover, the axisymmetric case is either a cigar or a pancake singularity, and the barrel singularity does not form.\footnote{For barrel singularity, $k_1=0, k_2, k_3>0$. These values of the Kasner exponents do not satisfy the constraint for the vacuum Bianchi-I given by eqs.\ (\ref{kasner1}) therefore barrel singularity does not form for dust and radiation filled Bianchi-I spacetime.} In the asymptotic limit when $a\rightarrow \infty$, it is the energy density which is more prominent and contribution of the anisotropic shear to the extrinsic curvature become negligible compared to that of the energy density. Hence the isotropization takes place. This is an important distinction from $w=1$ where the universe never isotropizes, not even in the asymptotic limit. It can also be concluded from the eq.\ (\ref{w01}) that smaller the equation of state, faster will be the isotropization. 
We now discuss some of the main features of the dust filled Bianchi-I universe. (The features for the radiation filled case can be deduced similarly). The behavior of the directional scale factors for a dust filled universe ($w=0$), also called the Heckmann-Shucking solution, is given as \cite{hs}
\be
\label{dusta}a_i(t) = a_{i\,o}\, t^{k_i}(t+t_o)^{2/3-k_i}
\ee
where $t_o$ is the initial time, and $k_i$ are the vacuum Kasner exponents satisfying the eqs.\ (\ref{kasner1}). 
As the classical singularity is approached i.e. when $t \ll t_o$, then it is straightforward to see from the eq.\ (\ref{dusta}) that the directional scale factors behave as $a_i(t) \propto t^{k_i}$.\footnote{For $t \ll t_o$, $t+t_o \approx t_o$ which is a constant and therefore the time dependence comes from the $t^{k_i}$ part of the expression.}
This precisely is the Kasner vacuum solution to the Bianchi-I spacetime. This shows that, near singularity, $w=0$ universe behaves like vacuum solution to Bianchi-I.
Note that the Kasner solution is only a limit which is approached when $t\ll t_o$, the universe never actually becomes an exact Kasner vacuum universe.
On the other hand the asymptotic limit $t\gg t_o$ gives the following expression for the directional scale factors
$
a_i(t)_{\rm dust}|_{t\rightarrow \infty} \propto t^{3/2}
$
and in this limit the isotropization takes place. Following similar argument for a radiation filled universe, one finds that 
the spacetime behaves like vacuum Kasner universe as the singularity is approached while in the asymptotic limit the spacetime isotropizes and the directional scale factors behave as
$
a_i(t)_{\rm rad}|_{t\rightarrow \infty} \propto t^{1/2}.
$
It is worth noting that the asymptotic behavior of these scale factors for dust and radiation filled Bianchi-I spacetime is the same as that of the scale factor of an isotropic spacetime filled with dust and radiation respectively.

\section {Effective dynamics in LQC and Kasner transitions}

In the previous section, we discussed the way various geometric structures arise in the Bianchi-I spacetime with stiff matter, dust and radiation in the classical theory. In this section, we will examine the formation of these structures in LQC using effective dynamics. We will study the structure of the spatial geometry on both sides of the bounce by computing the Kasner exponents and investigate the various possible transitions as the spacetime transits from an expanding to a contracting branch during the backward evolution of given initial data. The dependence of these transitions on the matter content and the anisotropy will also be explored in detail. We will, first, discuss the effective dynamics of Bianchi-I spacetime, and then in the following subsections we will present the results of numerical solution of the dynamical equations for stiff matter, dust and radiation filled universes.

\subsection{Effective dynamics}
The loop quantization of Bianchi-I is performed by expressing the classical Hamiltonian in terms of the elementary variables of LQG, namely the holonomies of the connection $A_a^i$ taken over the closed loops and the fluxes of the triads $E_i^a$ and promoting them to their corresponding quantum operators.\ This results in discrete quantum difference equations which govern the evolution of the spacetime.\ The discrete difference equations hence obtained are manifestation of the discrete quantum geometry inherent in loop quantization.
LQC also provides an approximate effective description of the underlying quantum geometry of the spacetime, known as the effective Hamiltonian description.\ The effective description is obtained through a faithful embedding of the finite dimensional classical phase space in the infinite dimensional quantum phase space of LQC based on the geometric formulation of quantum mechanics  \cite{schilling1, schilling2}.\ Through a proper choice of semiclassical states, this procedure furnishes an effective Hamiltonian which contains the leading order corrections due to the discrete quantum geometry. Using the Hamiltonian hence obtained, the effective equations of motion can be derived. These modified equations of motion introduce quantum geometric corrections to the classical dynamics of the spacetime and yield the modified dynamical equations.
 The dynamics hence obtained has been shown to be in excellent agreement with the full quantum evolution of  semiclassical states using numerical simulations in many situations including in the presence of anisotropies  \cite{aps2,aps3,bp,ap,apsv,kv,szulc_b1,madridb1}. Our analysis will assume validity of the effective Hamiltonian approach for the Bianchi-I model.

 The effective Hamiltonian constraint for Bianchi-I with choice of lapse $N=1$ can be written as \cite{awe2,csv,cv} \footnote{The effective Hamiltonian considered here does not include `inverse volume corrections' which are sometimes included in studies of spatially compact models in LQC. In our analysis  the Bianchi-I model is considered with a non-compact topology. Even if one takes a compact model, since the physical volume the fiducial cells in the simulations presented here are larger than Planck volume, the corrections introduced due to inverse triad terms would remain negligible.}
 \be
\label{b1heff} 
\Heff \, = {\mathcal H}_{\rm grav}+\Hmatt = \, -\f{1}{8 \pi G \gamma^2 V}\left(\sinamu \sinbmu p_1 p_2 +\mbox{cyclic terms}\right) \, + \, \Hmatt ~
 \ee
 where $\Hmatt$ is the matter Hamiltonian and ${\bar \mu}_i$ are given via
 \be
\bar\mu_1=\lambda \sqrt{\f{p_1}{p_2 p_3}}; \quad \bar\mu_2=\lambda \sqrt{\f{p_2}{p_1 p_3}} \quad {\rm and} \quad \bar\mu_3=\lambda \sqrt{\f{p_3}{p_1 p_2}} ~.
 \ee
 Here $\lambda^2=4\sqrt{3}\pi \gamma \lp^2$ arise from the discrete quantum geometry of the underlying spacetime predicted by LQG. The dependence of $\bar \mu_i$ on the $p_i$ is unique in the sense that any other choice of the function $\bar \mu_i (p_i)$ would lead the physical results to depend on the rescaling of the edges of the fiducial cell and the shape of the fiducial cell would affect the physical results \cite{cs09}. 
  
  The equations of motion of the triads and connection can be computed using the Hamilton's equations of motion-
  
  \be
  \label{p1d}\dot{p_1}\,=\,\f{p_1}{\gamma \lambda} \left(\sinb+\sinc \right)\cosa
  \ee
 
 and 
 \ba
\label{c1d} {\dot c}_1\, & =& \nonumber \, \f{1}{p_1\gamma\lambda}\bigg[c_2p_2\cos\left(\mub c_2\right)\left(\sin(\mua c_1)+\sin(\muc c_3)\right)\,+\,c_3p_3\cos\left(\muc c_3\right)\left(\sin(\mua c_1)+\sin(\mub c_2)\right)\, \\ && \nonumber-\, c_1p_1\cos\left(\mua c_1\right)\left(\sin(\mub c_2)+\sin(\muc c_3)\right) \, -\, \f{2 \mua p_2p_3}{\lambda} \bigg(\sin(\mub c_2)\sin(\muc c_3)\, \\ && +\, \sin(\muc c_3)\sin(\mua c_1)\, +\, \sin(\mua c_1)\sin(\mub c_2) \bigg)\bigg] + 8 \pi G \gamma \f{\partial \Hmatt}{\partial p_1}~.
 \ea
 Similarly one can obtain equations for $\dot p_2,\, \dot p_3,\, \dot c_2\, {\rm and}\, \dot c_3$, and also $\ddot a_i/a_i$ \cite{ps11}.  
 
 The values of the directional triads and the connection at any time can be computed through numerical evolution of initial data specified at a given instant of time. The directional Hubble rates, the expansion scalar and the shear scalar can be computed using the definitions given in eqs.\ (\ref{h1def}, \ref{thetadef} and \ref{sheardef}) respectively.
The dynamical equations of  motion for $p_i\, {\rm and}\, c_i$ (given in eqs.\ \ref{p1d} and \ref{c1d} and their cyclic permutation) modify the Einstein's field equations. From the eq.\ (\ref{p1d}) it is evident that since the maximum value of $\sin(\theta)$ is $1$, the value of $\dot p_i/p_i$ never diverges and hence the modified effective equations provide a singularity free solutions, causing bounces of the directional scale factors (or the bounce of the mean scale factor). The energy density, expansion scalar and the shear scalar are also bounded by their respective upper maxima \cite{gs1}. Further, all strong singularities for the matter with equation of state $w > -1$ have been shown to be resolved \cite{ps11}. 

Before we go in to detailed analysis for different matter components, it is important to make following remarks.\\

\noindent
{\bf Remark 1:} In the classical theory, depending on the matter content and the anisotropy, Bianchi-I spacetime demonstrates various geometrical structures of singularity; for example if one of the three directional scale factors diverges and the other two approach to zero as the singularity is approached, then the structure of the spatial geometry of the spacetime tends to an `infinite' cigar. In LQC, for the same initial data, the structure near the bounce is rather a `finite' cigar. This is so because in LQC, due to the nonsingular bounce the singularity is avoided and all the scale factors take non-zero and finite value near the bounce. Similarly, if the classical trajectory for a given initial data collapses to a barrel, point or pancake singularity, then in LQC, these geometric structures form for the same initial data, but are finite. Although LQC and GR trajectories are in agreement with each other in the low curvature regime and an observer sees both trajectories tending towards the same type of geometrical structure as the high curvature regime is approached, the actual spatial geometry near the bounce in LQC is non-vanishing and is a finite version of the geometry in GR near singularity. \\

\noindent
{\bf Remark 2:} In the following analysis we always specify the initial data in the expanding branch (post-bounce) of the mean scale factor and perform a backward time evolution through the bounce to the contracting branch (pre-bounce). A Kasner transition in the following is defined as the change in geometric structure from the post-bounce phase of the mean scale factor $a$, as observed by an observer approaching the bounce, to the one in the pre-bounce phase as will be observed by the same observer if she/he evolves forward in time in the contracting branch. Thus, even though the evolution occurs in only one direction (i.e.\ backward), the geometric structure in the pre- and post-bounce phases is defined in the approach to the bounce. Note that the nature of transition is unaffected by providing the initial data in the contracting branch of the mean scale factor and evolving in future time direction.

 \subsection{Bianchi-I LQC with stiff matter}

\begin{figure}[tbh!]
\includegraphics[angle=0,width=0.47\textwidth]{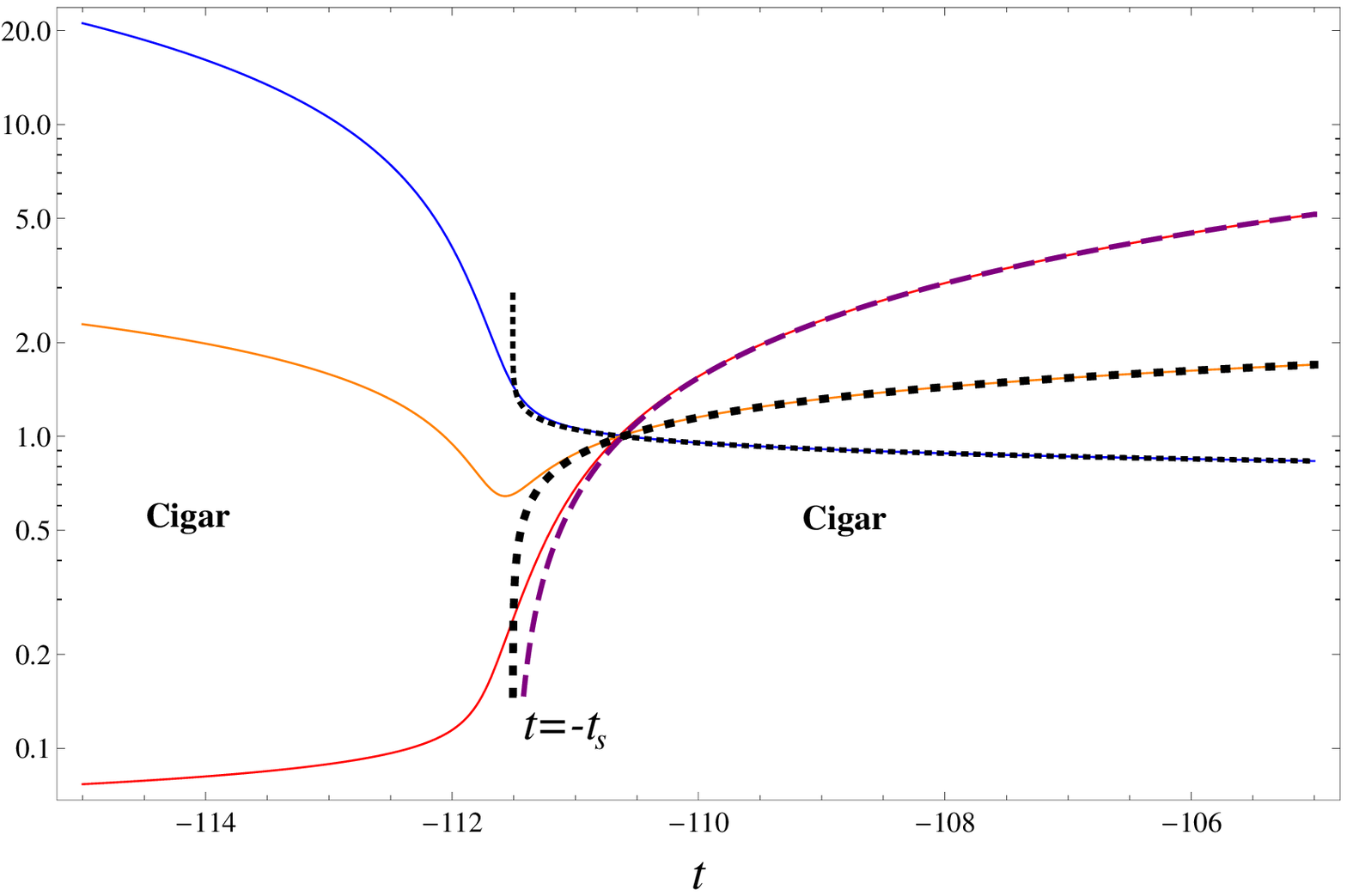}
\hskip0.5cm
\includegraphics[angle=0,width=0.47\textwidth]{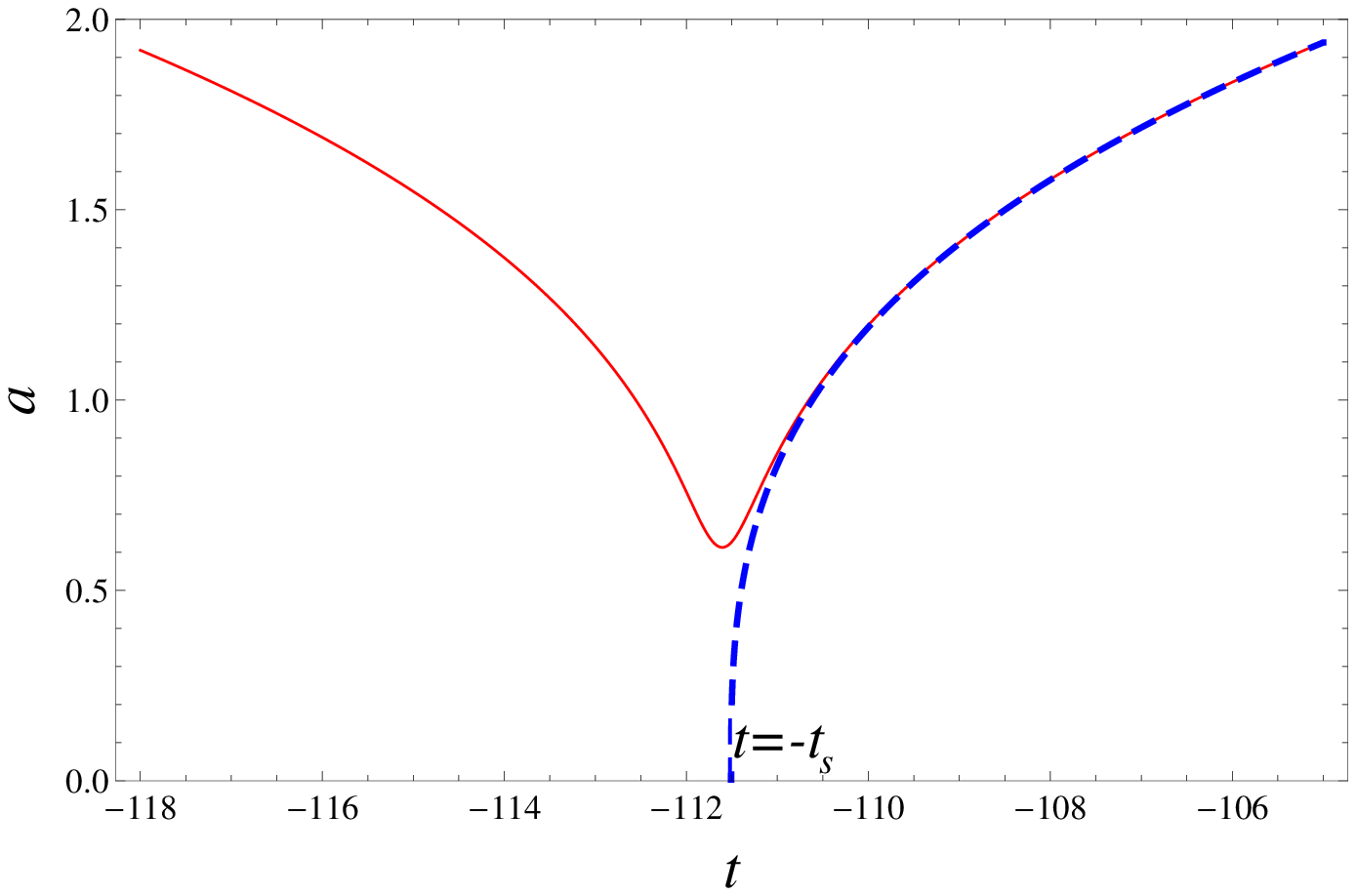}
\caption{The left plot shows the cigar-cigar transition of structure of the spatial geometry of the spacetime for $1/\sqrt{3}\, < \,  \delta \, < \, 1$, the dotted lines represent the corresponding classical trajectory and the right plot shows the evolution of the mean scale factor. The spacetime undergoes a cigar-cigar transition while the mean scale factor bounces and remains non-singular throughout the evolution.}
\label{a123}
\end{figure}

We perform several numerical simulations (over 100 simulations for each range of $\delta$) for various anisotropy parameters to solve the system of differential equations (\ref{p1d} and \ref{c1d}) for a variety of initial data. We use 4th order adaptive Runge-Kutta for the numerical solutions to the equations. Then, for various ranges of the anisotropy parameters we find out the structure of spacetime by calculating the values of the Kasner exponents. 
This way a wide class of transition of structures is obtained, based on the initial anisotropy.
It also turns out that the structures of spacetime on two sides of the bounce are not the same although the anisotropic shear parameter $\Sigma^2$ is conserved on the two sides of the bounce in the asymptotic limit. This provides a rich and interesting transitions of structures across the bounce. \fref{a123} shows an example of a Kasner transition across the non-singular bounce. We find that in the post-bounce phase, two scale factors are decreasing and one is increasing as the bounce of the mean scale factor is approached, i.e.\ a cigar is formed. Similarly, in the pre-bounce phase, two scale factors decrease while one increases, on approach to the bounce in the forward time evolution. Thus, again a finite cigar forms. Hence, this figure corresponds to a cigar-cigar transition. (In all the plots, presented in the following, structures are defined in this same way).

\begin{figure}[tbh!]
\includegraphics[angle=0,width=0.47\textwidth]{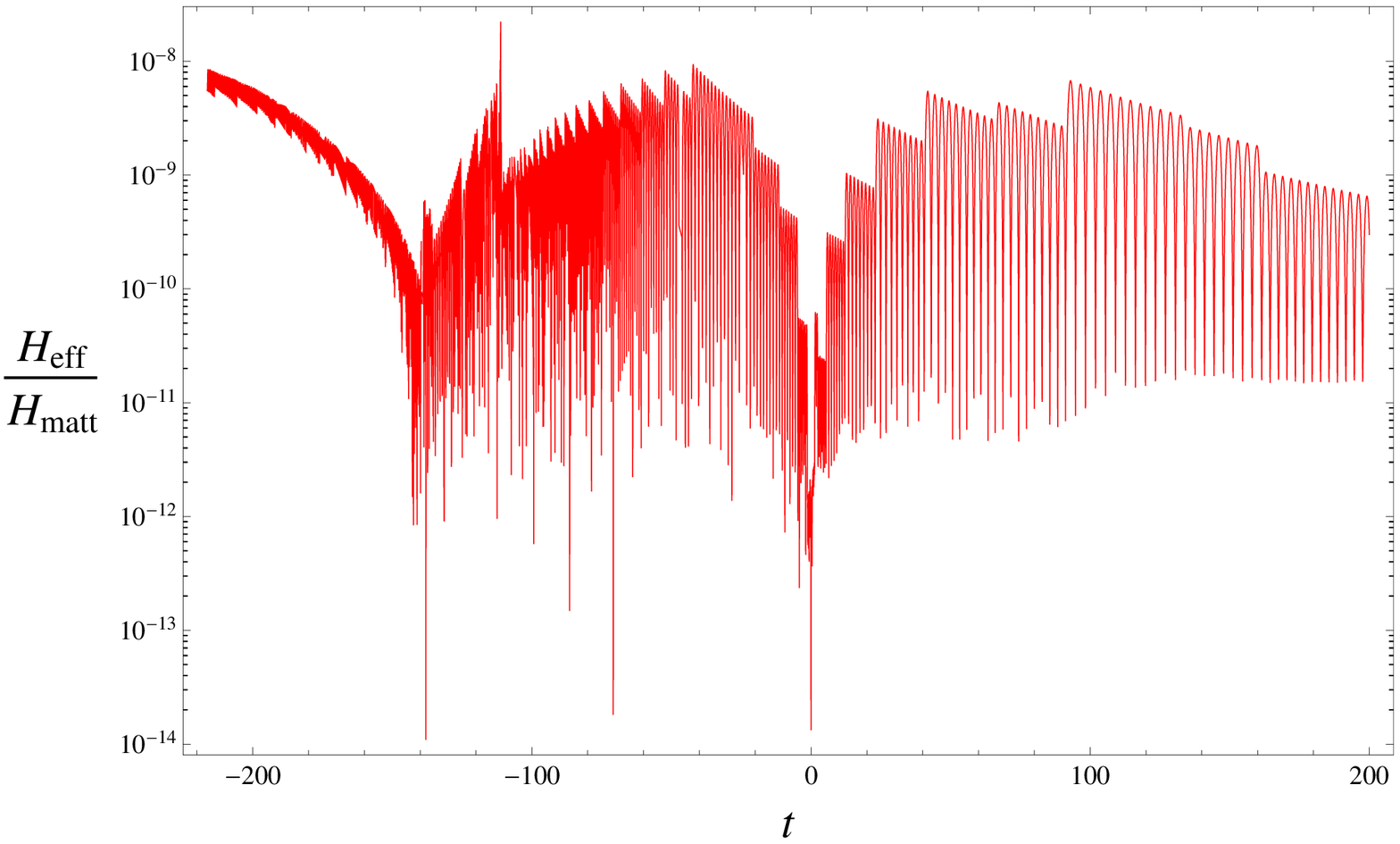}
\hskip0.5cm
\includegraphics[angle=0,width=0.47\textwidth]{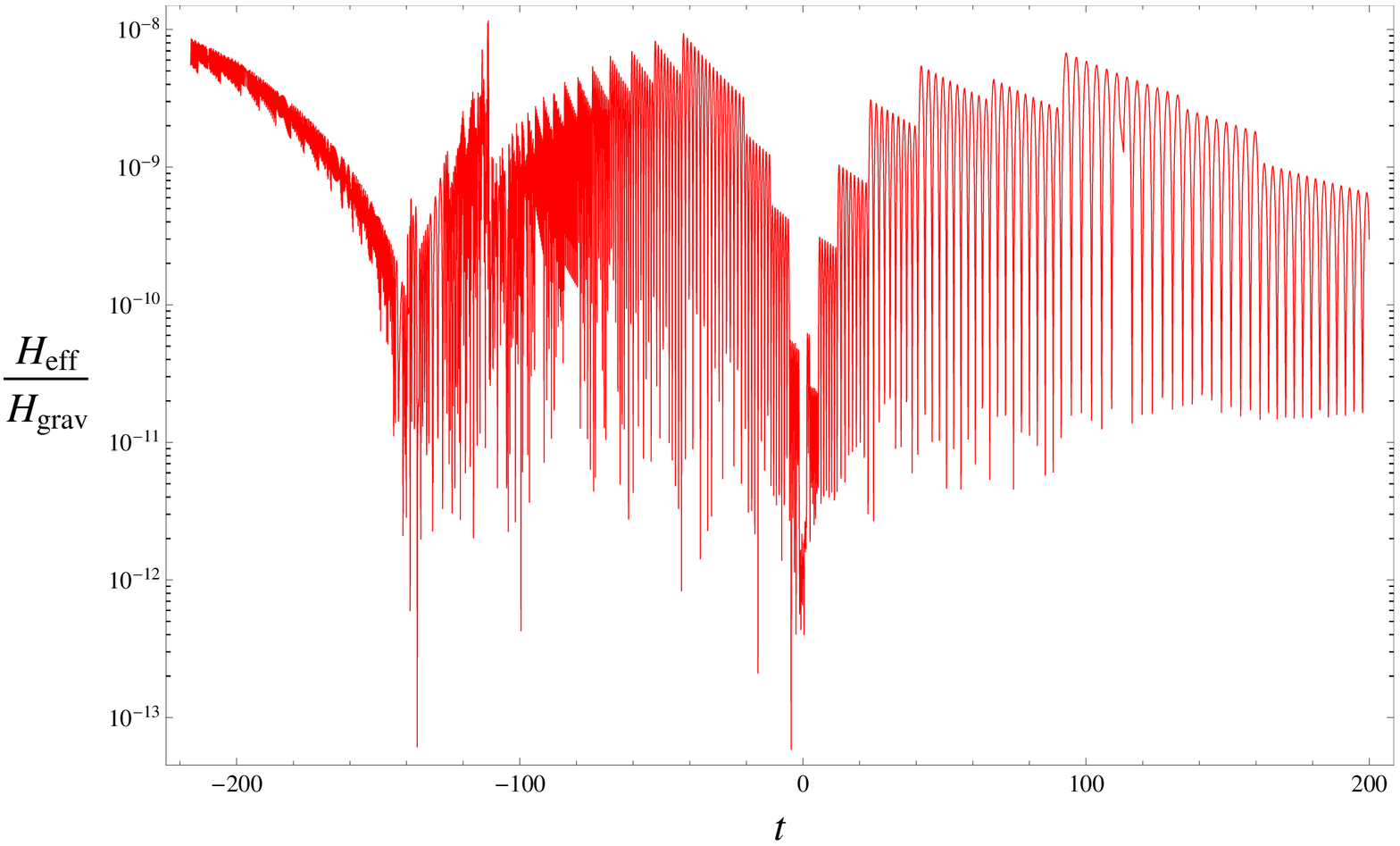}
\caption{This plot shows the Hamiltonian constraint during the evolution of the effective equations. $|{\mathcal H}_{\rm eff}/\Hmatt|$ and $|\Heff/{\mathcal H}_{\rm grav}|$ remain sufficiently small at all the time which indicates the high numerical accuracy of the results.}
\label{hamilt}
\end{figure}

In all numerical simulations presented here, the initial data is given at $t=0$ without any loss of generality and the classical singularity occurs at $t=-t_s$. The value of $t_s$ depends on the initial data. The initial data set consists of the values of the connections ($c_1, c_2$), the triads ($p_1,p_2,p_3$), the energy density $\rho$ at time $t=0$. The value of the remaining connection component $c_3$ is computed by satisfying the effective Hamiltonian constraint given by eq.\ (\ref{b1heff}). The parameters $\psi$ and $\delta$ are computed via eq.\ (\ref{psidef} and \ref{deldef}) by using the Kasner exponents and the directional scale factors, which can be easily obtained from the triads $p_i$. This provides the complete specification of the initial condition of the system. Also, the constant $\psi_c$ (which will be used further in this section) is given in terms of the anisotropy parameter ($\delta$) as $\psi_c=\sin^{-1}({|\delta|}/{2})$.

\fref{hamilt} shows an important feature of the numerical simulations, the absolute value of the ratios of effective Hamiltonian to the matter Hamiltonian and gravitational part of the Hamiltonian are of the order of $10^{-8}$ or less during the evolution, which implies that the effective Hamiltonian constraint is satisfied with such a high order of accuracy. This, in turn, assures the numerical accuracy of the results presented in this article up to the order of $10^{-8}$. We consider such a numerical accuracy to be good enough for our analysis because the physical quantities of interest are of  order much higher than $10^{-8}$ and hence the numerical errors can safely be neglected.

 We now discuss the cases of various values of the anisotropy parameter $\delta$.

\subsubsection{$1/\sqrt{3}\, < \,  \delta \, < \, 1$}

This range of anisotropy parameter $\delta$ corresponds to high anisotropy and it is the only range in positive $\delta$ region when cigar to cigar transition is allowed. All possible transitions depending on $\psi$ are shown in the Table-I. Out of the total range of $\psi=\left[0, 2\pi/3\right]$, cigar to cigar, of all the transitions, occupies the biggest range. As we will see in the following subsections that cigar-cigar transition is forbidden for $|\delta| \leq 1/\sqrt{3}$. This signals that cigar-cigar transition is a highly anisotropic transition. Axisymmetric solutions exist for $\psi=\left( \f{2\pi}{3}-\psi_c\, {\rm and}\, \f{\pi}{3}+\psi_c\right)$ on the expanding side of the universe and for $\psi= \left(\f{\pi}{3}-\psi_c\, {\rm and}\, \psi_c\right)$ on the contracting side. As discussed in  Sec. II, the parameter $\delta$ is related to the ratio of the shear scalar to the energy density of the matter field and larger value of $\delta$ means that the anisotropy dominates over the matter energy density. This is partially the reason why there are no point like to point like transitions allowed for this range of $\delta$.

\begin{figure}[tbh!]
\includegraphics[angle=0,width=0.45\textwidth]{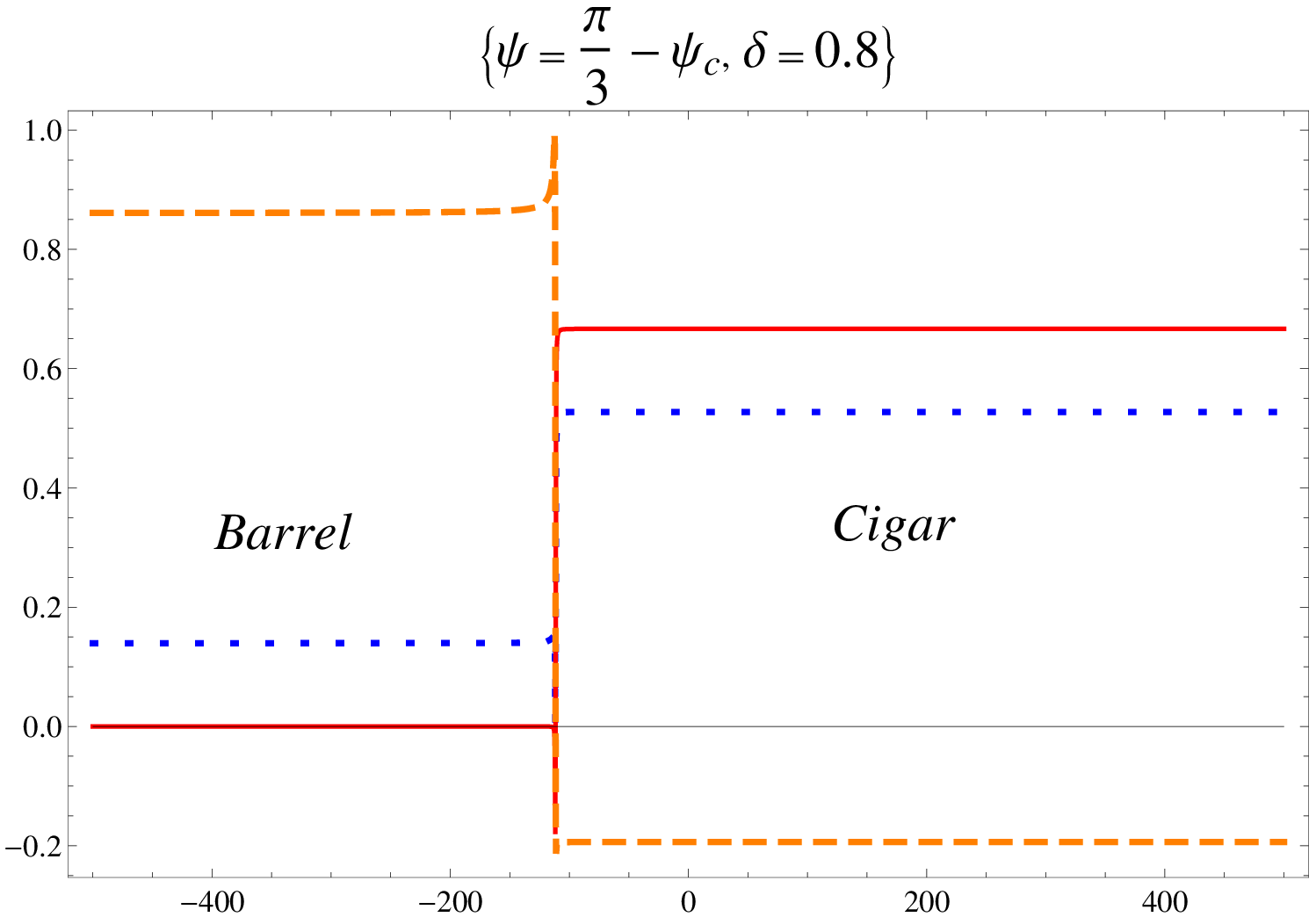}
\hskip0.5cm
\includegraphics[angle=0,width=0.45\textwidth]{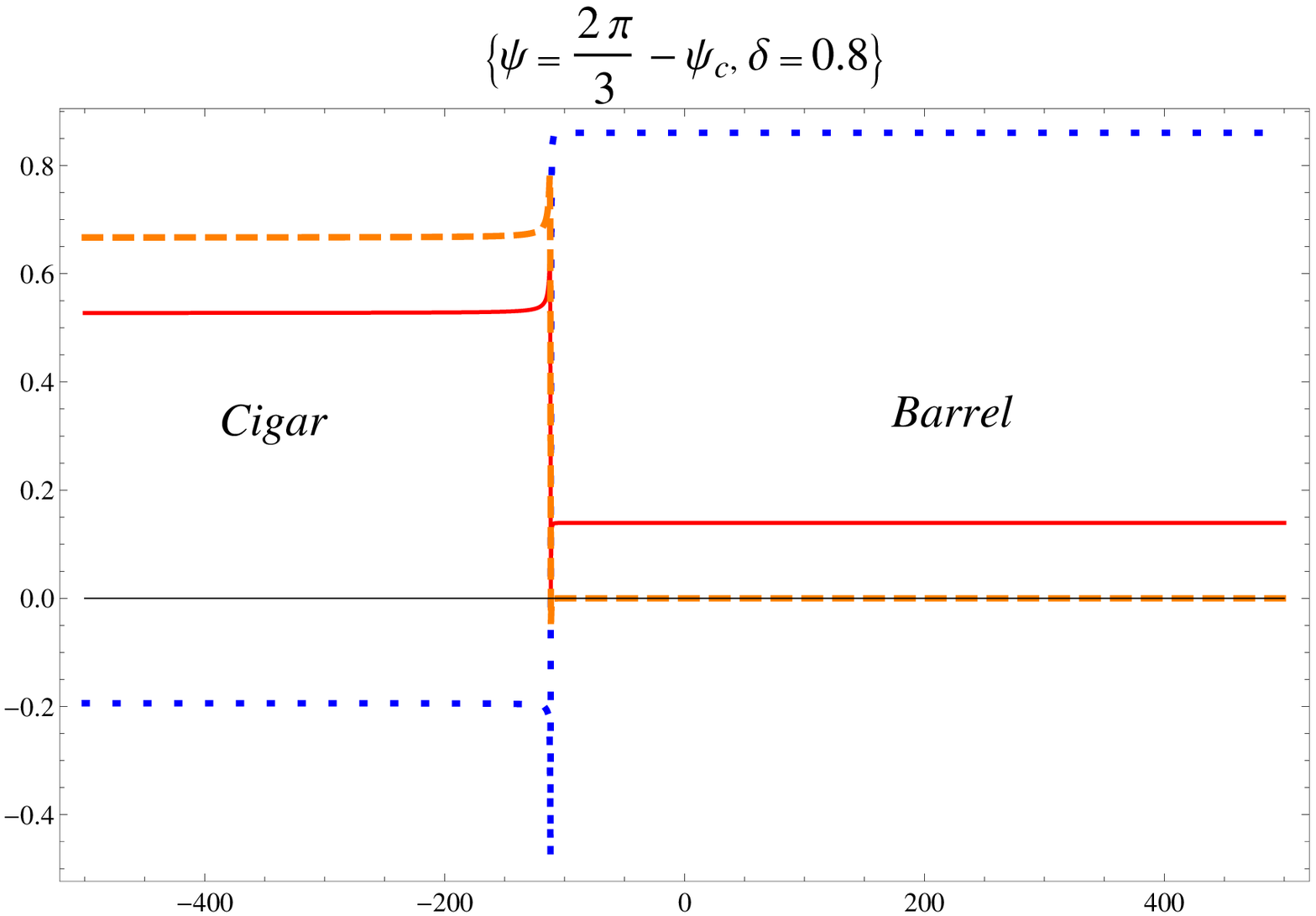}
\vskip0.5cm
\includegraphics[angle=0,width=0.45\textwidth]{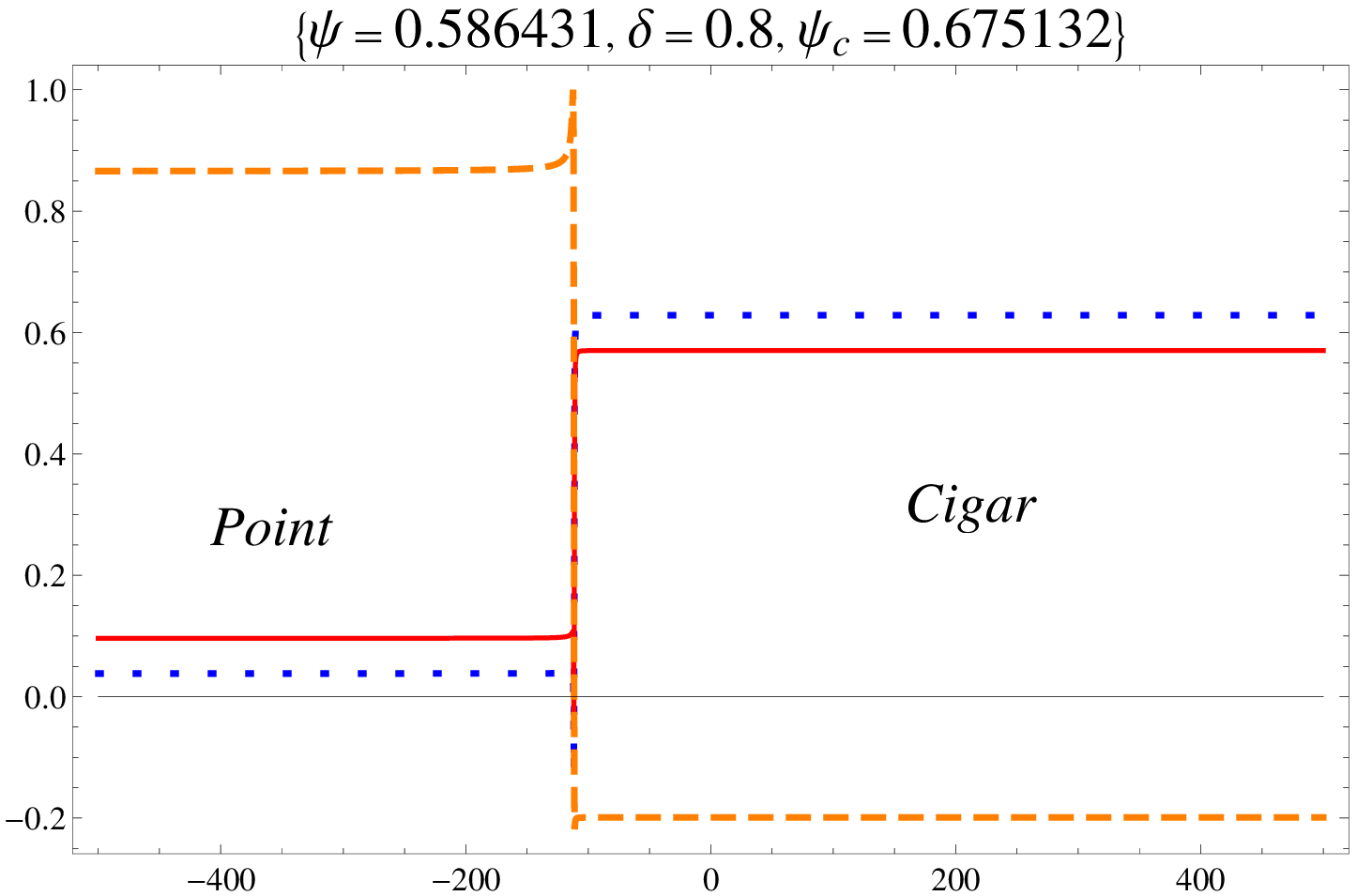}
\hskip0.5cm
\includegraphics[angle=0,width=0.45\textwidth]{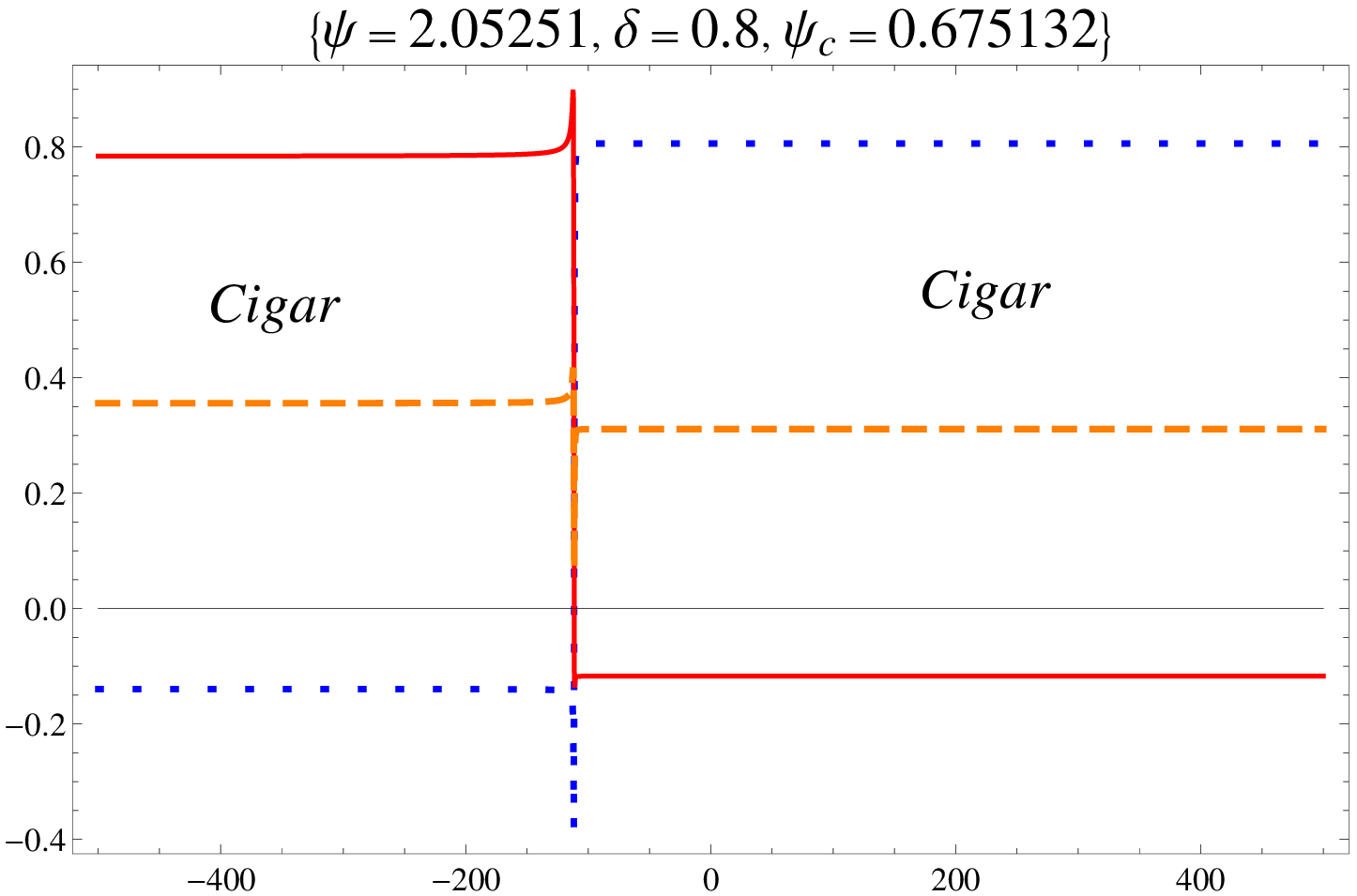}
\caption{These plots show various transitions of structure across the bounce depending on various values of $\psi$ for $1/\sqrt{3}\, < \,  \delta \, < \, 1$.}
\label{del08}
\end{figure}

\begin{figure}[tbh!]
\includegraphics[angle=0,width=0.45\textwidth]{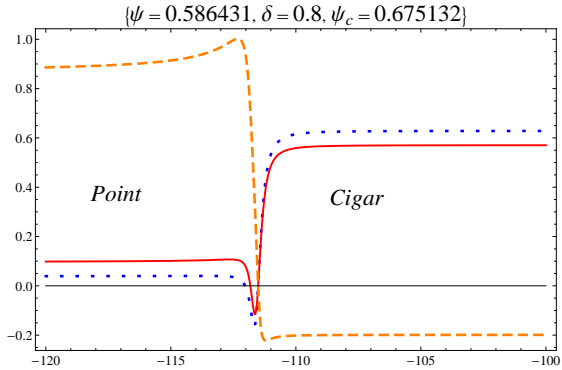}
\caption{This figure shows the transition of Kasner exponents (from cigar to point like) zoomed in near the bounce for $\delta=0.8$ and $\psi=0.58543$. }
\label{del08zoom}
\end{figure}

  Various possible transitions for a choice of $\delta=0.8$ are shown in \fref{del08}. The transition for $\delta=0.8$ and $\psi=0.586431$ (which correspond to cigar to point like transition) zoomed in near the bounce has been shown in \fref{del08zoom} which clearly portrays the rich dynamics near the bounce and resulting transition in the Kasner exponents. It is straightforward to see that the values of the Kasner components change across the bounce leading to the corresponding transition in the structure of the spatial geometry. Similar to the Kasner transitions in classical mixmaster phenomena, one of the Kasner components retains its behavior and the remaining two switch their role or two of them retain their behavior while the remaining one changes its sign. In other words, if $k_1<0\, {\rm and }\, k_2,k_3 >0$ (cigar) in the expanding branch, then after the transition takes place the Kasner components are either  $k_1,k_2>0\, {\rm and}\, k_3<0$ (as shown in the cigar-cigar transition where one of the Kasner exponents remains positive on both sides while the remaining two change their sign) or $k_1>0\, {\rm and }\, k_2,k_3 >0$ (as shown for the cigar to point like transition where only of the exponents changes the sign while the other tow remain positive across the bounce). The table also shows that if there is an axisymmetric barrel structure on one side, it goes to cigar type structure across the bounce. There are no barrel to barrel, or barrel to point like transitions allowed for this range of $\delta$. The transitions involving barrel happen for a certain value of $\psi$, whereas other transitions have a range of $\psi$ associated to them.

\subsubsection{$|\delta|=1/\sqrt{3}$}
\begin{figure}[tbh!]
\includegraphics[angle=0,width=0.45\textwidth]{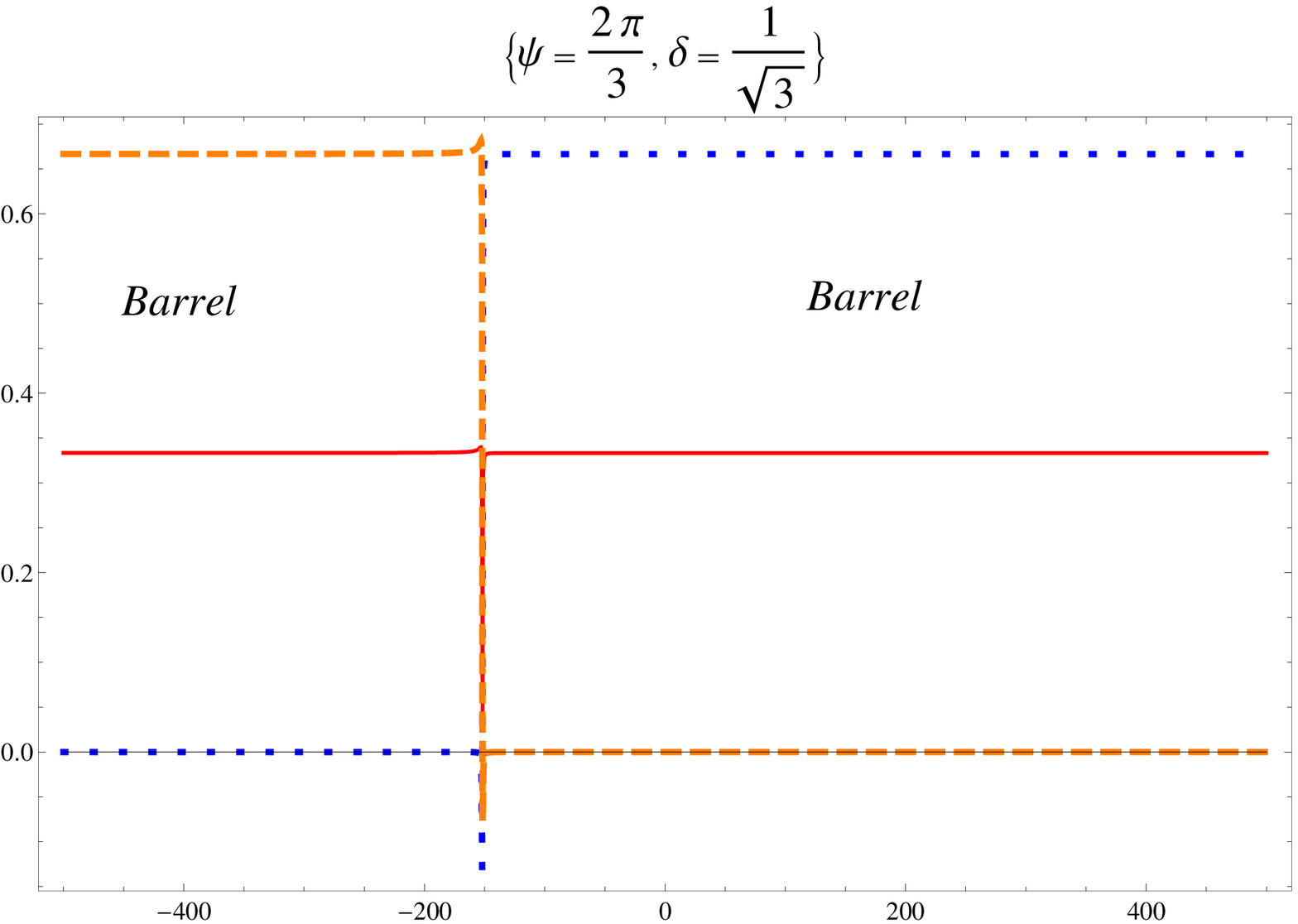}
\hskip0.5cm
\includegraphics[angle=0,width=0.45\textwidth]{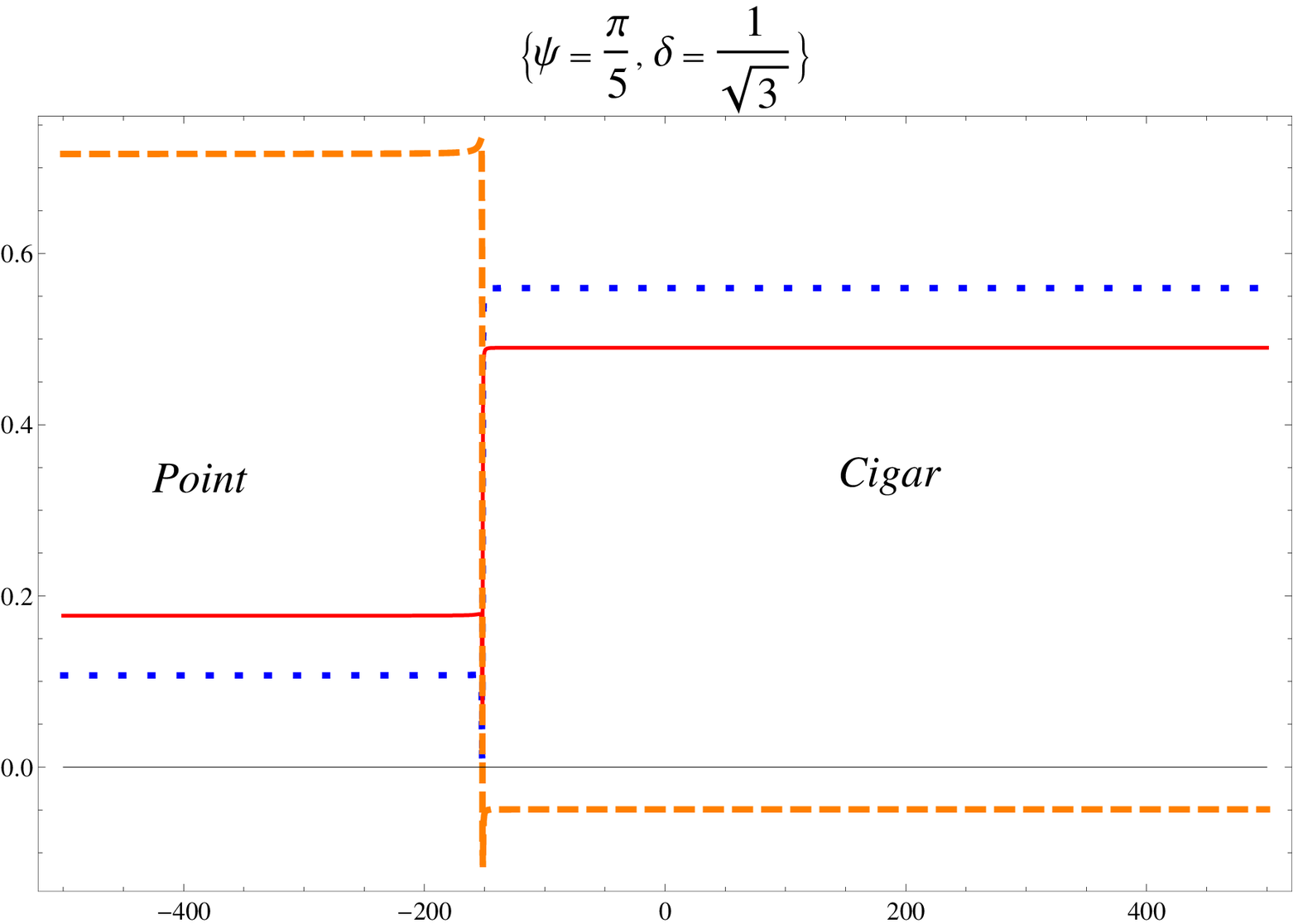}
\vskip0.5cm
\includegraphics[angle=0,width=0.45\textwidth]{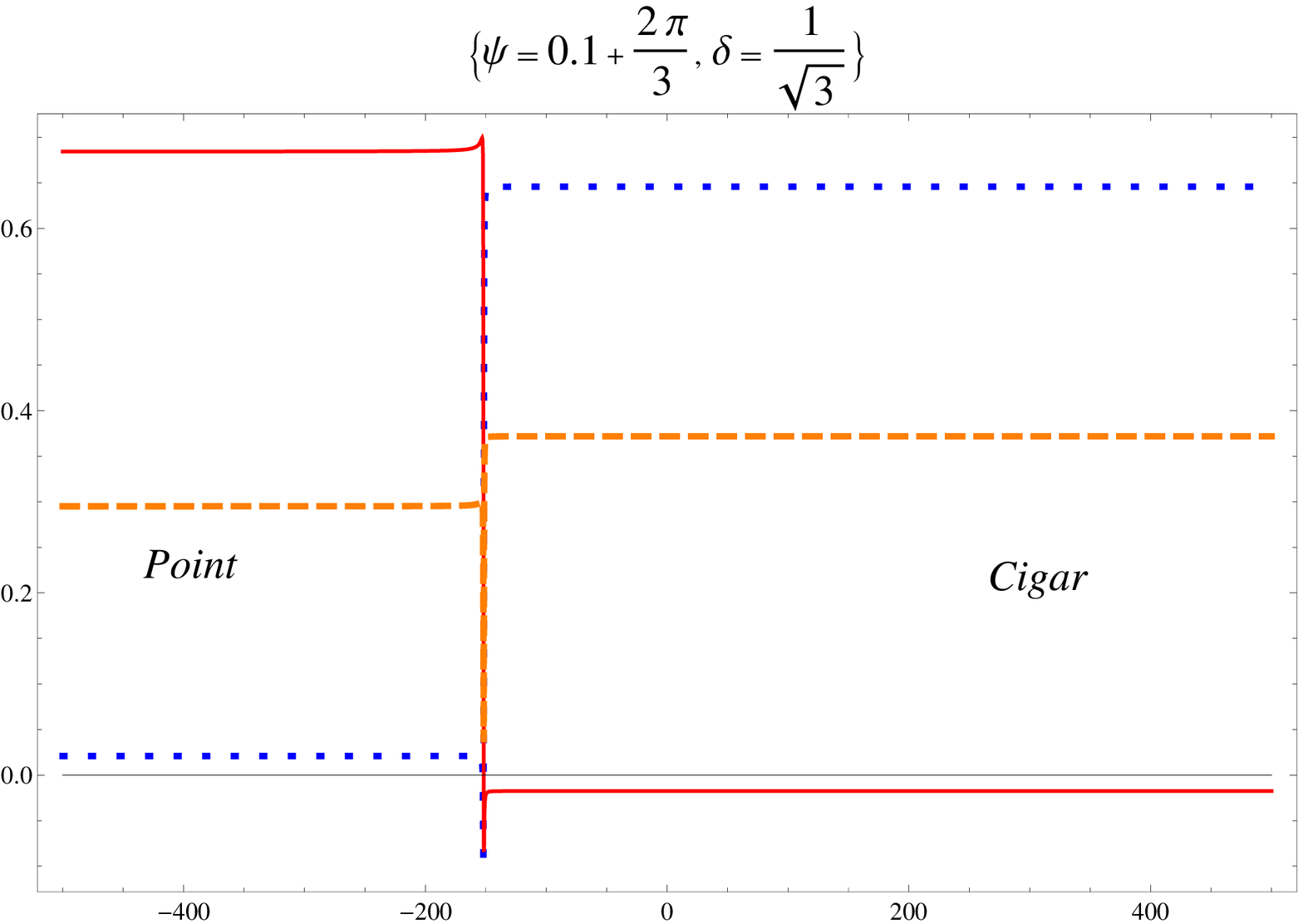}
\hskip0.5cm
\includegraphics[angle=0,width=0.45\textwidth]{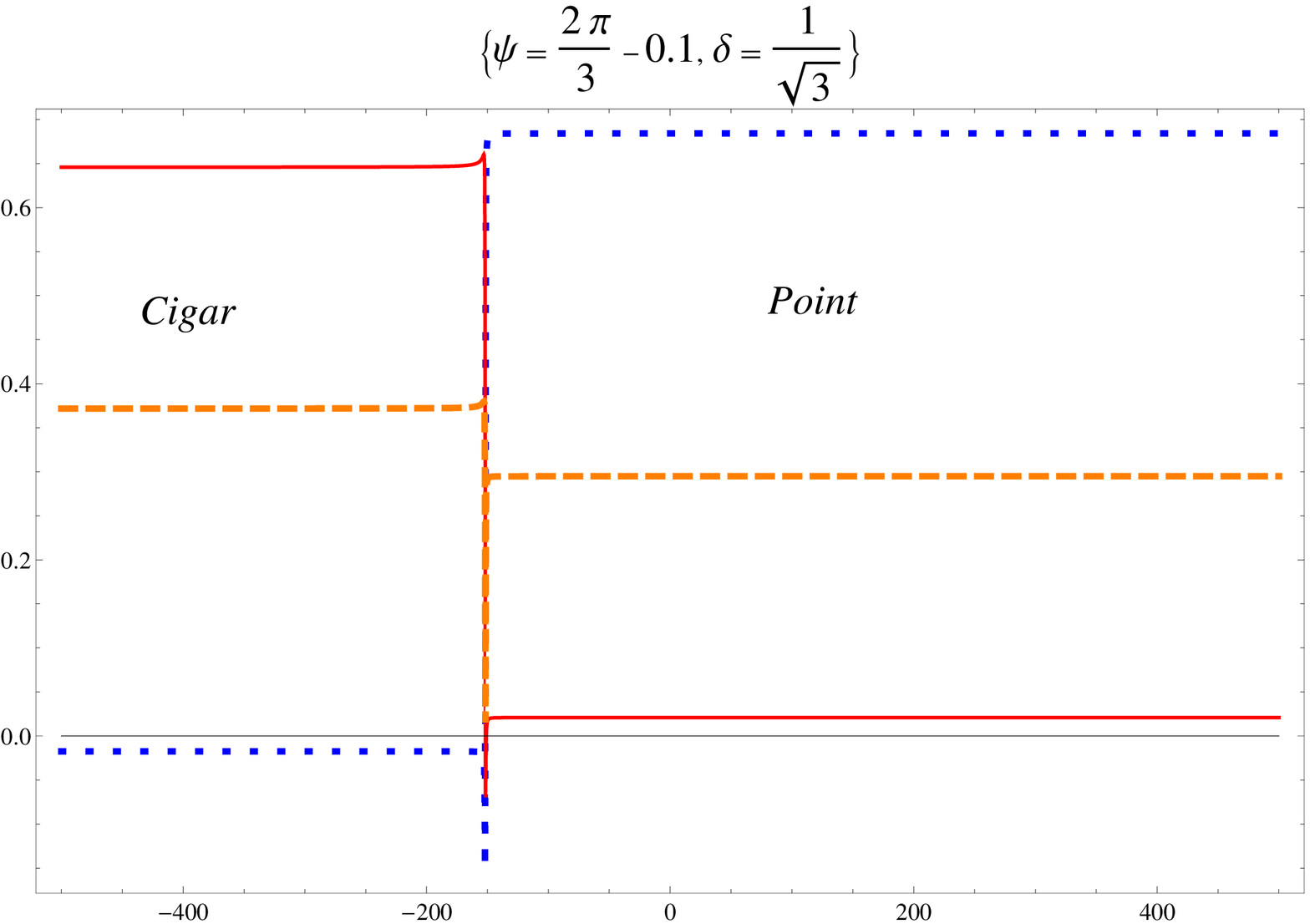}
\caption{Various transitions of structure across the bounce for $\delta=1/\sqrt{3}$. The Barrel-Barrel transition occurs at $\psi=2\pi/3$, a small deviation from $\psi=2\pi/3$ results in other transitions for example when $\psi$ decreased, the transition is Point-Cigar and when increased, the transition is Cigar-Point.}
\label{1sqrt3}
\end{figure}

\begin{figure}[tbh!]
\includegraphics[angle=0,width=0.45\textwidth]{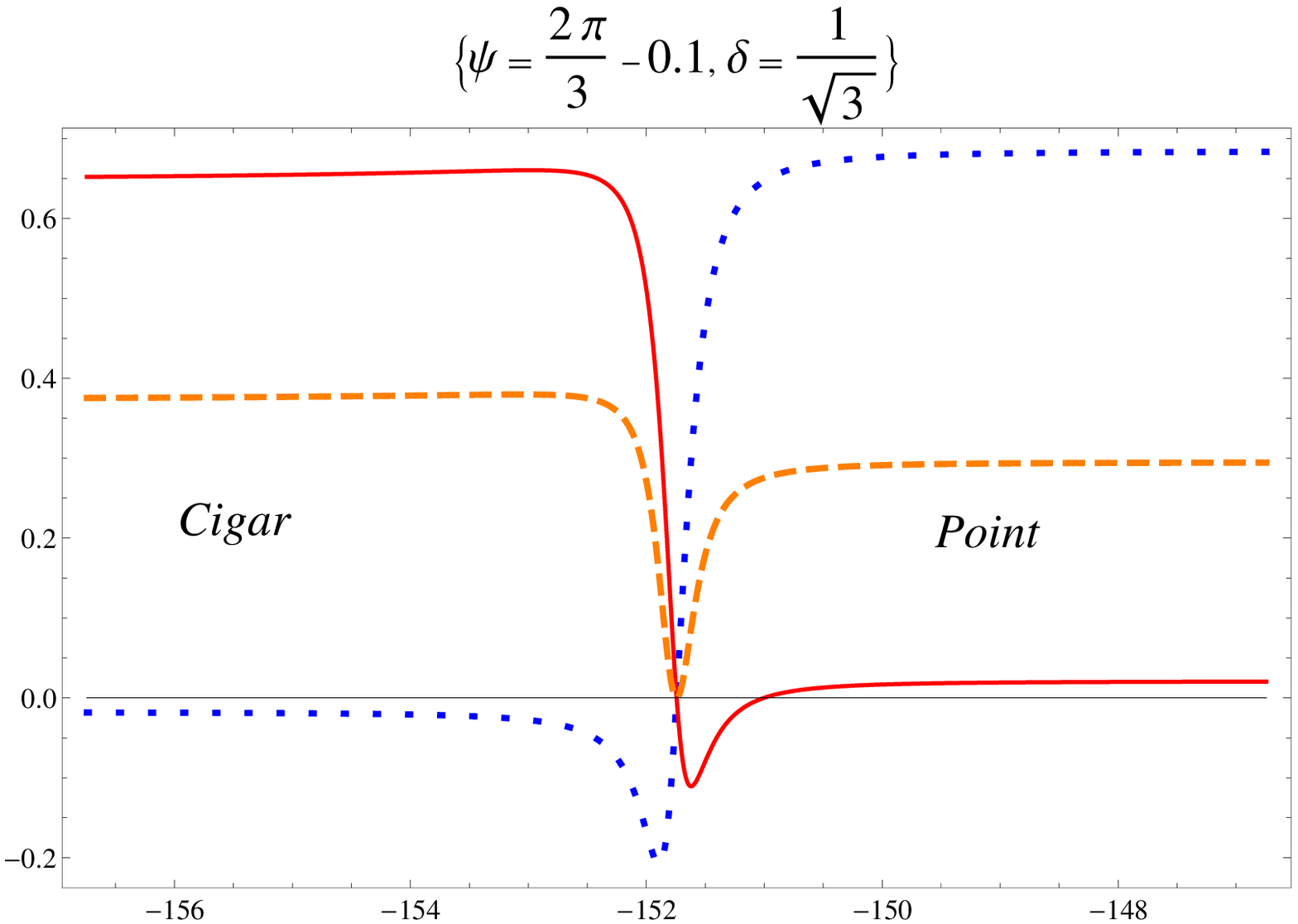}
\caption{This figure shows the transition of Kasner exponents (from point like to cigar) zoomed in near the bounce for $\delta=1/\sqrt{3}$ and $\psi=2\pi/3 -0.1$. }
\label{1sq3zoom}
\end{figure}

$\delta=1/\sqrt{3}$ is very interesting case in the sense that this is the only case where the barrel to barrel transition occurs. In all the other cases if there is barrel structure on one side of the bounce, it becomes either cigar or point like on the other side. Except for the two values of $\psi$ when the barrel structure is allowed, for all the other $\psi$ the transitions are the point like to cigar transitions. Moreover, the cigar-cigar transition is forbidden in this range. The barrel to barrel transition occurs at particular values of $\psi$ unlike for other transitions for which there are ranges of $\psi$. 
\fref{1sqrt3} shows the Kasner transition for $\delta=1/\sqrt{3}$.  \fref{1sq3zoom} shows one of the transitions of \fref{1sqrt3} ($\delta=1/\sqrt{3}, \psi=2\pi/3-0.1$) zoomed in near the bounce. In a barrel-barrel transition one of the Kasner exponents remain the same while the other two interchange their value across the bounce.
It is important to note that the transitions involving barrel occur at a fixed value of $\psi$ for a given $\delta$ while other structures happen for a range of values of $\psi$. This feature is shown in \fref{1sqrt3} -- the barrel-barrel transition occurs at $\psi=2\pi/3$, and if $\psi$ is decreased by a small amount, the structures on both sides of the bounce change and the resulting transition becomes point like to cigar. Similarly, when $\psi$ increased slightly, cigar to point like transition turns up.

\begin{table}[tbh!]
	{\textbf{Table-I}} \\[3pt]
	\begin{tabular}{|c|c|c|}
	\hline
	&Expanding branch & Contracting branch\\[5pt]
	\hline
$1/\sqrt{3} <  \delta < 1$	&	Cigar                                                         &    Cigar $0 < \, \psi <\, \f{\pi}{3}-\psi_c$\\
						        &	$0 \, < \, \psi <\, \f{2\pi}{3}-\psi_c$            &    Barrel $\psi=\f{\pi}{3}-\psi_c$ \\
							&                                                                &    Point $\f{\pi}{3}-\psi_c <\, \psi < \, \psi_c$ \\
							&                                                                 &    Barrel $\psi=\psi_c$ \\
							&                                                                &    Cigar $\psi_c < \, \psi < \, \f{2\pi}{3}-\psi_c$\\\cline{2-3}
	
        						&	Barrel                                                        & Cigar \\
						&        $\psi=\f{2\pi}{3}-\psi_c$                            & $\psi=\f{2\pi}{3}-\psi_c$\\\cline{2-3}

						&	Point 							& Cigar \\
						 &	$\f{2\pi}{3}-\psi_c < \, \psi  < \, \f{\pi}{3}+\psi_c$ &$\f{2\pi}{3}-\psi_c < \, \psi  < \, \f{\pi}{3}+\psi_c$ \\\cline{2-3}

						&	Barrel 							& Cigar \\
						&	$\psi=\f{\pi}{3}+\psi_c$				& $\psi=\f{\pi}{3}+\psi_c$\\\cline{2-3}

						&        Cigar   							& Cigar \\
					        &	$\f{\pi}{3}+\psi_c < \, \psi < \, \f{2\pi}{3}$	& $\f{\pi}{3}+\psi_c < \, \psi < \, \f{2\pi}{3}$\\\cline{2-3}
	\hline
	
 $ \delta=1/\sqrt{3}$ 			&       Barrel \,  $\psi =0$                                                           &    Barrel\,  $\psi=0$ \\\cline{2-3}

						&        Cigar \, $0 < \,\psi < \, \f{\pi}{3}$                                      & Point \,  $0 < \,\psi < \, \f{\pi}{3}$\\\cline{2-3}

						&	Barrel \, $\psi =\f{\pi}{3}$  							& Barrel \, $\psi =\f{\pi}{3}$\\\cline{2-3}
							
						&	Point \, $\f{\pi}{3} < \,\psi < \, \f{2\pi}{3}$ 				& Cigar\, $\f{\pi}{3} < \,\psi < \, \f{2\pi}{3}$ \\

	\hline

 $1/2 < \delta < 1/\sqrt{3}$ &       Point                                                         &    Point \\
						&        $0 \, \leq \, \psi <\, \psi_c-\f{\pi}{3}$          	&    $0 \, \leq \, \psi <\, \psi_c-\f{\pi}{3}$ \\\cline{2-3}
	
						&        Barrel                                                        & Point \\
						&        $\psi=\psi_c-\f{\pi}{3}$                              & $\psi=\psi_c-\f{\pi}{3}$ \\\cline{2-3}

						&	Cigar 							                 & Point \\
				&	$\psi_c-\f{\pi}{3} < \, \psi < \, \f{2\pi}{3}-\psi_c $       & $\psi_c-\f{\pi}{3} < \, \psi < \, \f{2\pi}{3}-\psi_c $ \\\cline{2-3}

				&	Barrel 							& Point \\
				&	($\psi=\f{2\pi}{3}-\psi_c$)				&($\psi=\f{2\pi}{3}-\psi_c$)\\\cline{2-3}

 				&        Point   								& Point ($\f{2\pi}{3}-\psi_c <\psi <\psi_c$) \\
				&        $(\f{2\pi}{3}-\psi_c <\psi <\f{2\pi}{3})$		& Barrel ($\psi =\psi_c$)\\ 
        				&						& Cigar ($\psi_c <  \psi <  \pi-\psi_c$) \\
				&						& Barrel ($\psi=\pi-\psi_c$ ) \\
				&						& Point  ($\pi-\psi_c < \psi < \f{2\pi}{3}$)\\
	\hline
	\end{tabular}
	\caption{Transition of structures across the bounce for ($1/2\leq\delta<1$)}
	\label{table1}
\end{table}

\subsubsection{$1/2<\delta<1/\sqrt{3}$}
\begin{figure}[tbh!]
\includegraphics[angle=0,width=0.45\textwidth]{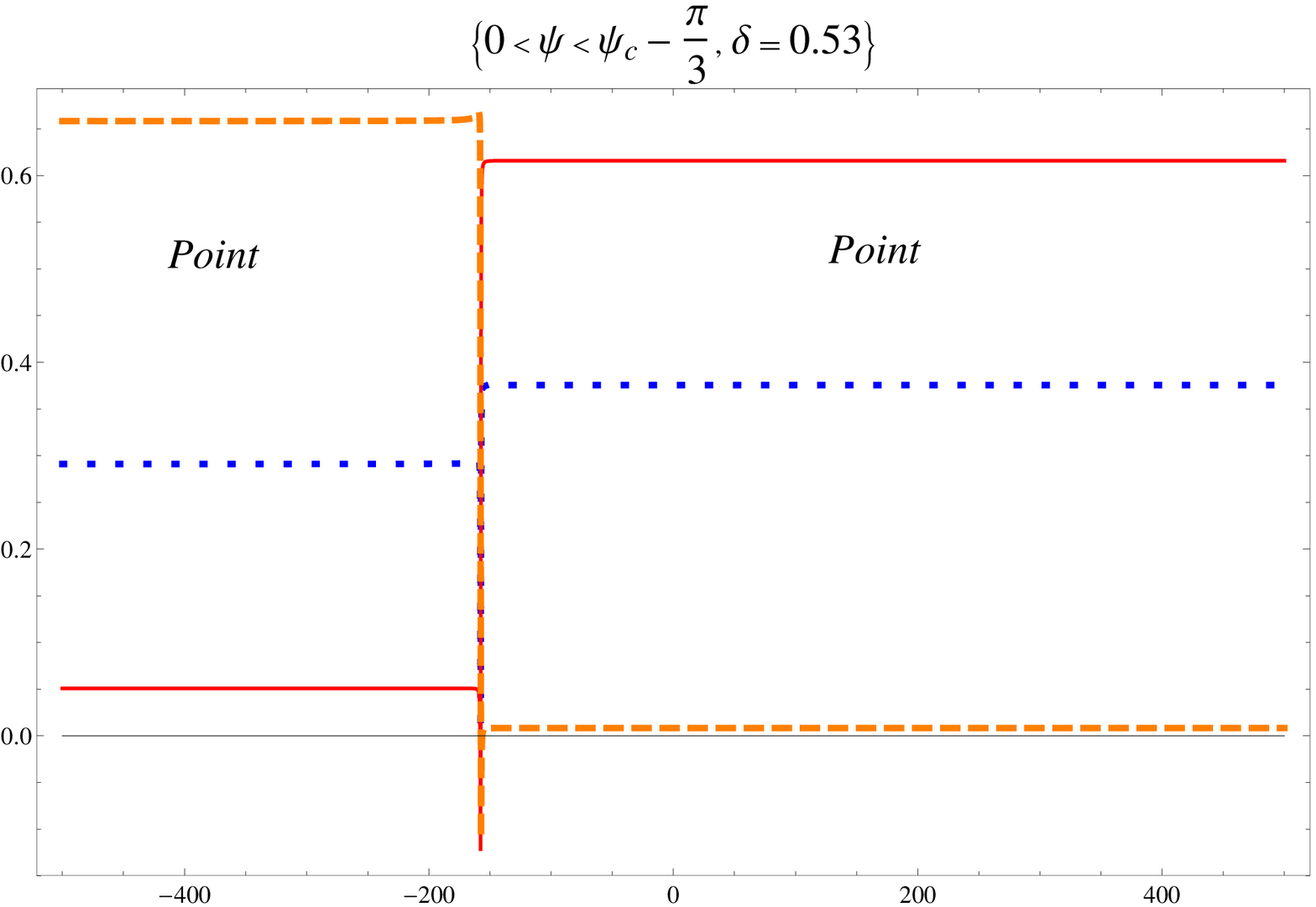}
\hskip0.5cm
\includegraphics[angle=0,width=0.45\textwidth]{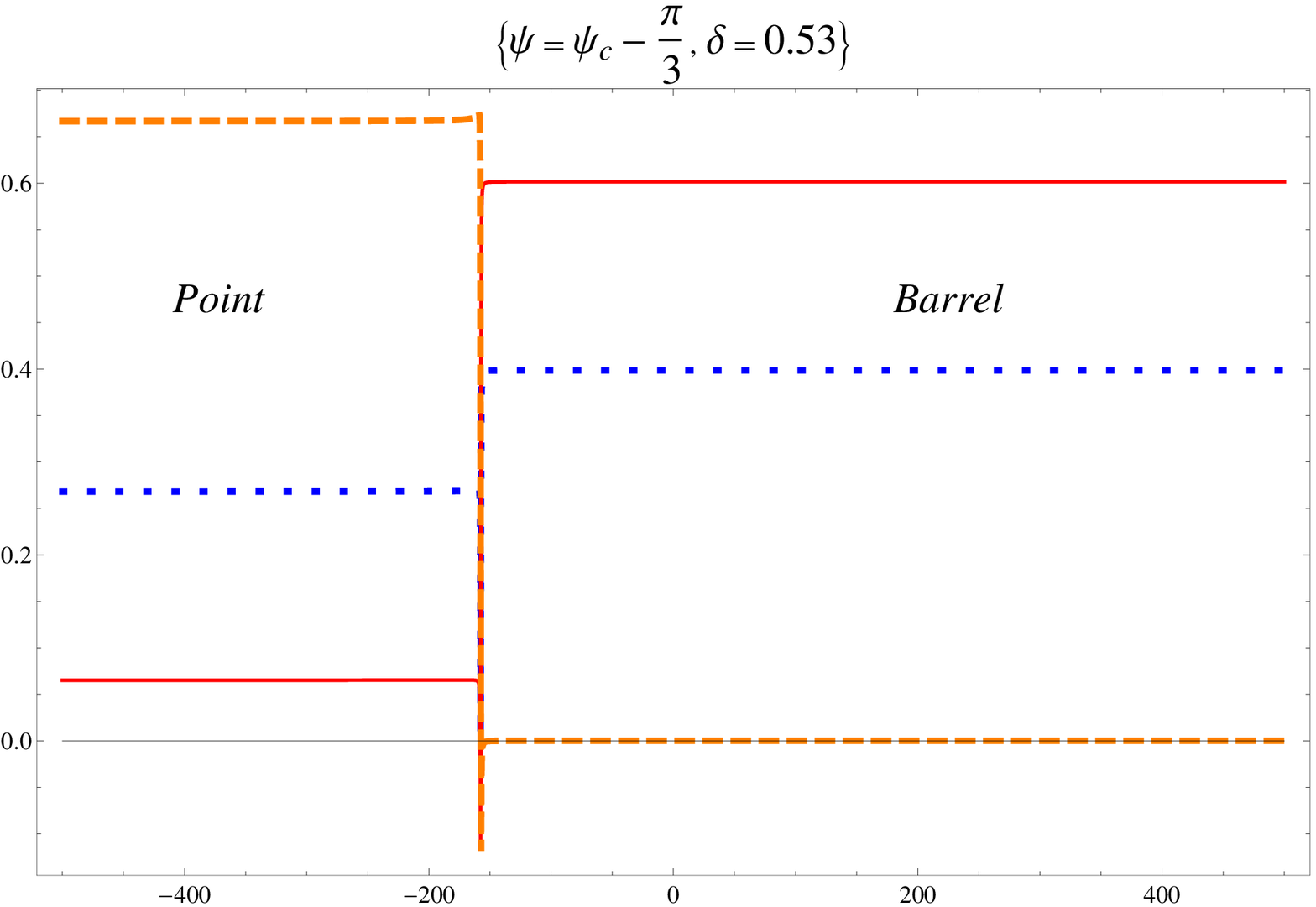}
\vskip0.5cm
\includegraphics[angle=0,width=0.45\textwidth]{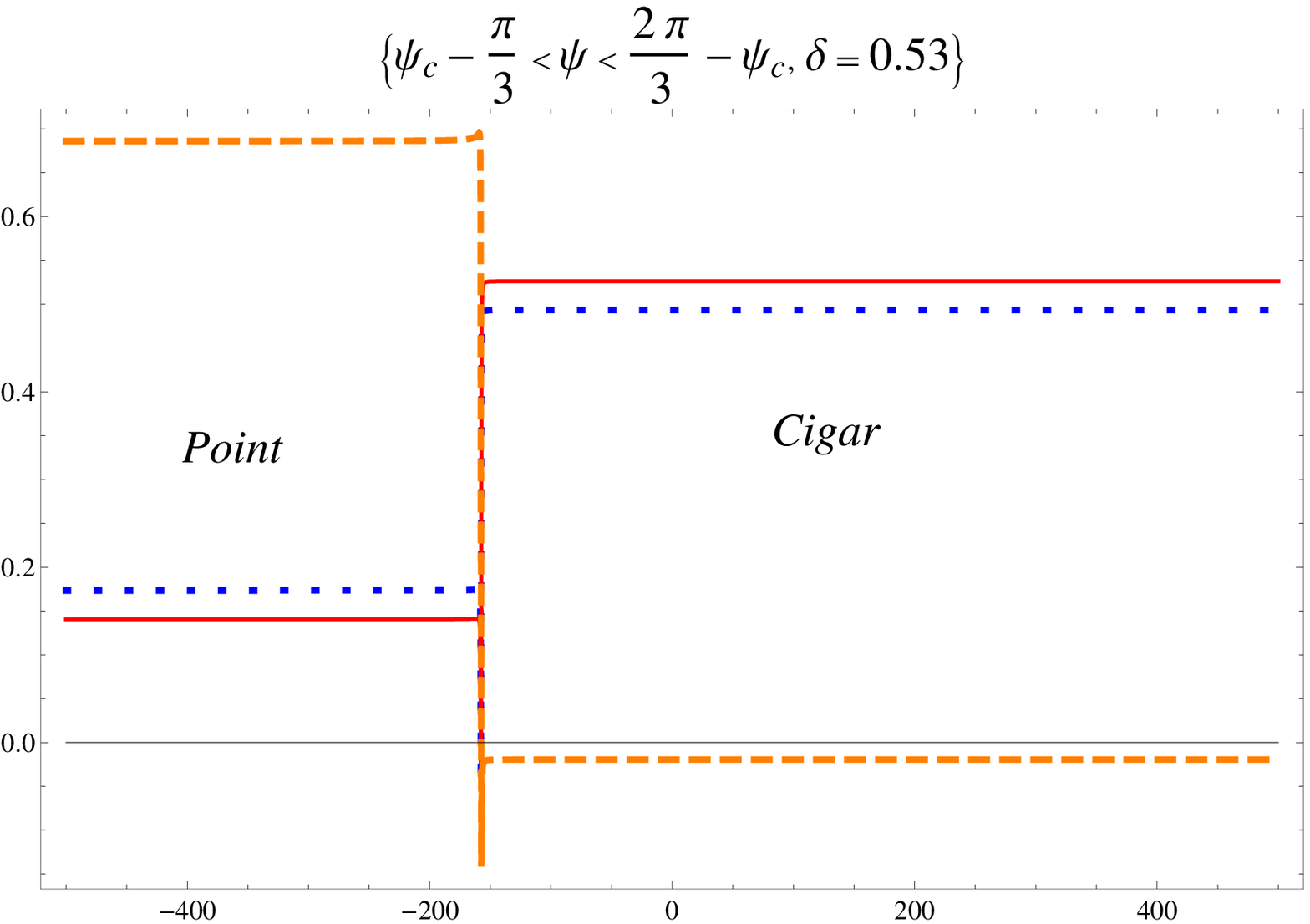}
\hskip0.5cm
\includegraphics[angle=0,width=0.45\textwidth]{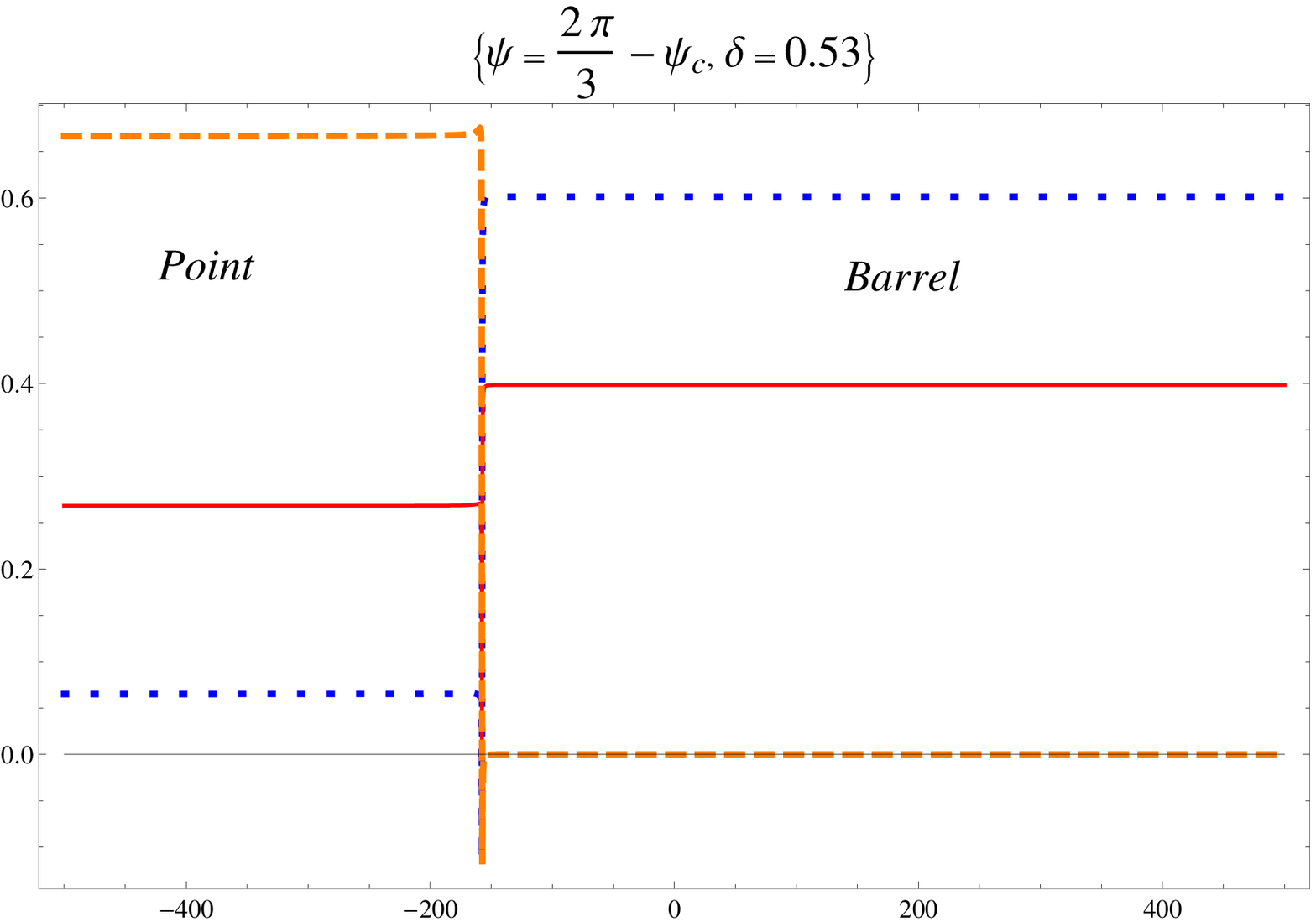}
\caption{This figure shows the transition of structure across the bounce for $1/2<\delta<1/\sqrt{3}$ when $\delta=0.53$.}
\label{del053}
\end{figure}

\begin{figure}[tbh!]
\includegraphics[angle=0,width=0.45\textwidth]{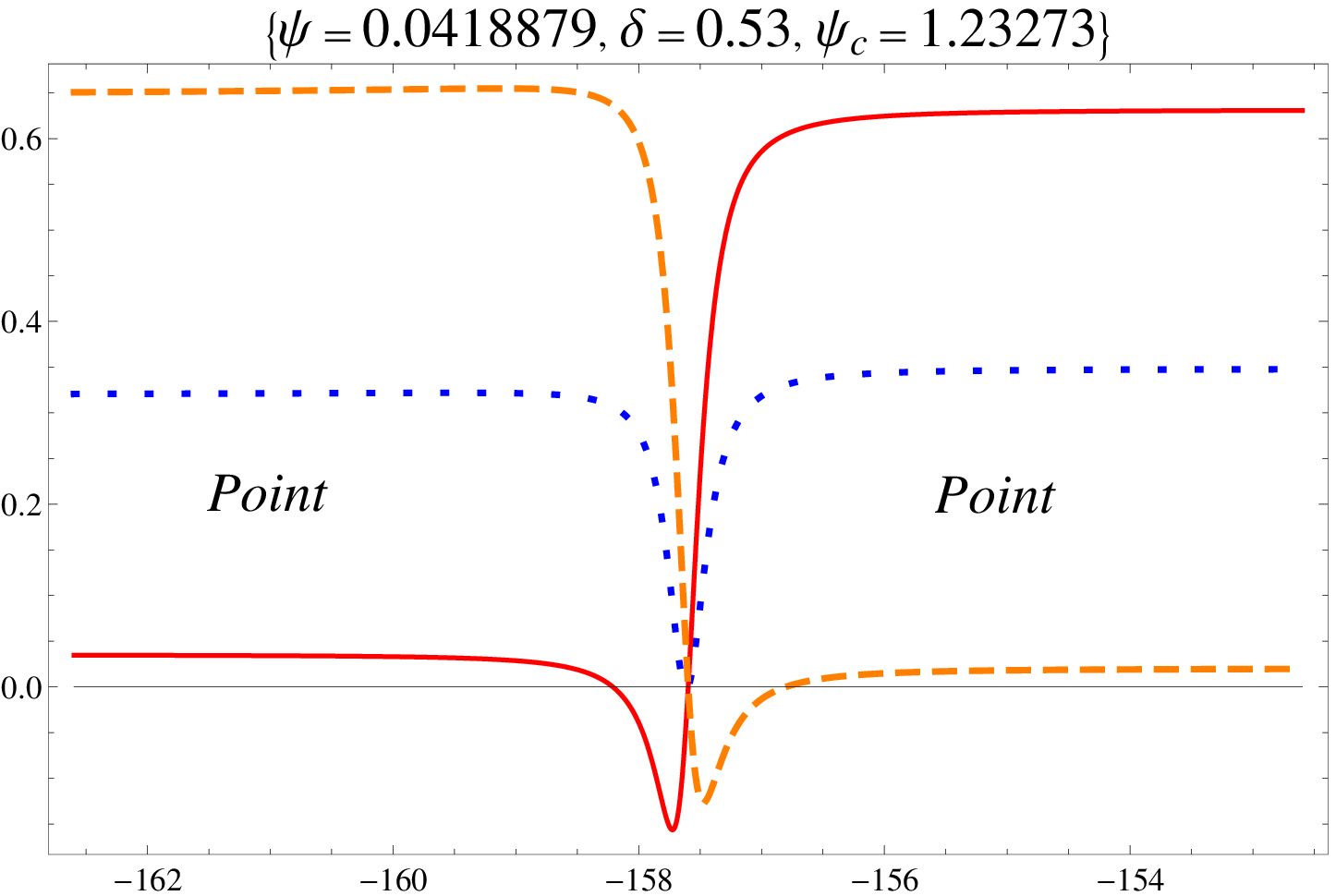}
\caption{This figure shows the transition of Kasner exponents (from point like to point like) zoomed in near the bounce for $\delta=0.53$ and $\psi=0.04188$. }
\label{del53zoom}
\end{figure}

In this range of $\delta$, point like to point like transition appears but there are no cigar-cigar transitions anymore. Also the barrel-barrel transition are now forbidden. The absence of cigar-cigar transitions and the appearance of point like to point like transition are consistent with the fact that the anisotropy is now smaller than the previous cases, and the matter dominates over the anisotropic shear. \fref{del053} shows  various transitions for $\delta=0.53$. One of the transitions ($\delta=0.53, \psi=0.04188$), which correspond to point like to point like transition, has been zoomed in near the bounce in \fref{del53zoom}. It is worth noting that barrel to point like transition occurs at a fixed value of $\psi=\psi_c-\pi/3$ and if $\psi$ is changed even for a small amount, the transition of structures changes.  As shown in figure if $\psi$ is decreased, the transition is point like to cigar and when increased it becomes cigar to point like. Again, at $\psi=2\pi/3-\psi_c$ the transition involves the barrel structure in the expanding branch.

\subsubsection{$|\delta|\leq 1/2$}
\begin{figure}[tbh!]
\includegraphics[angle=0,width=0.45\textwidth]{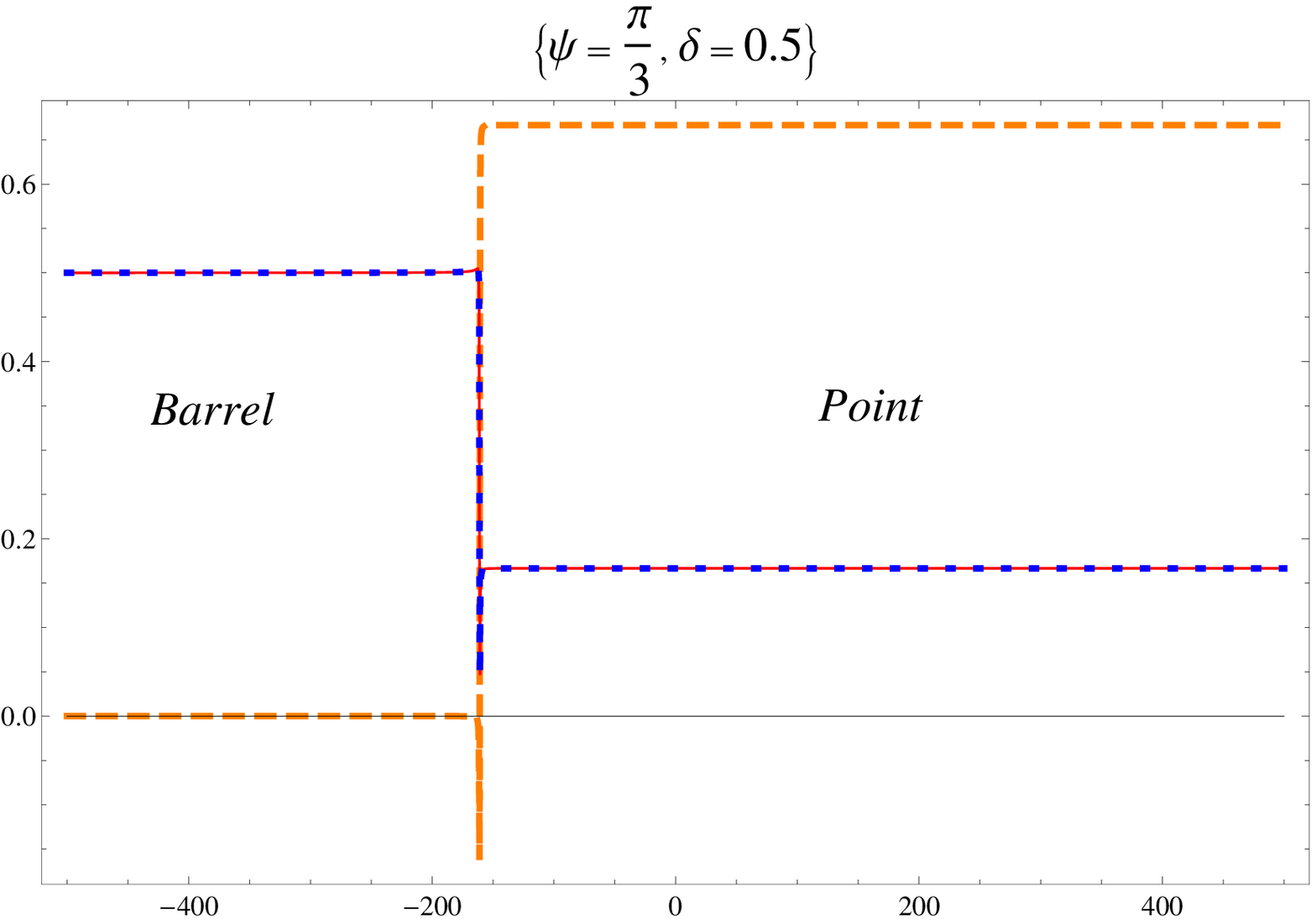}
\hskip0.5cm
\includegraphics[angle=0,width=0.45\textwidth]{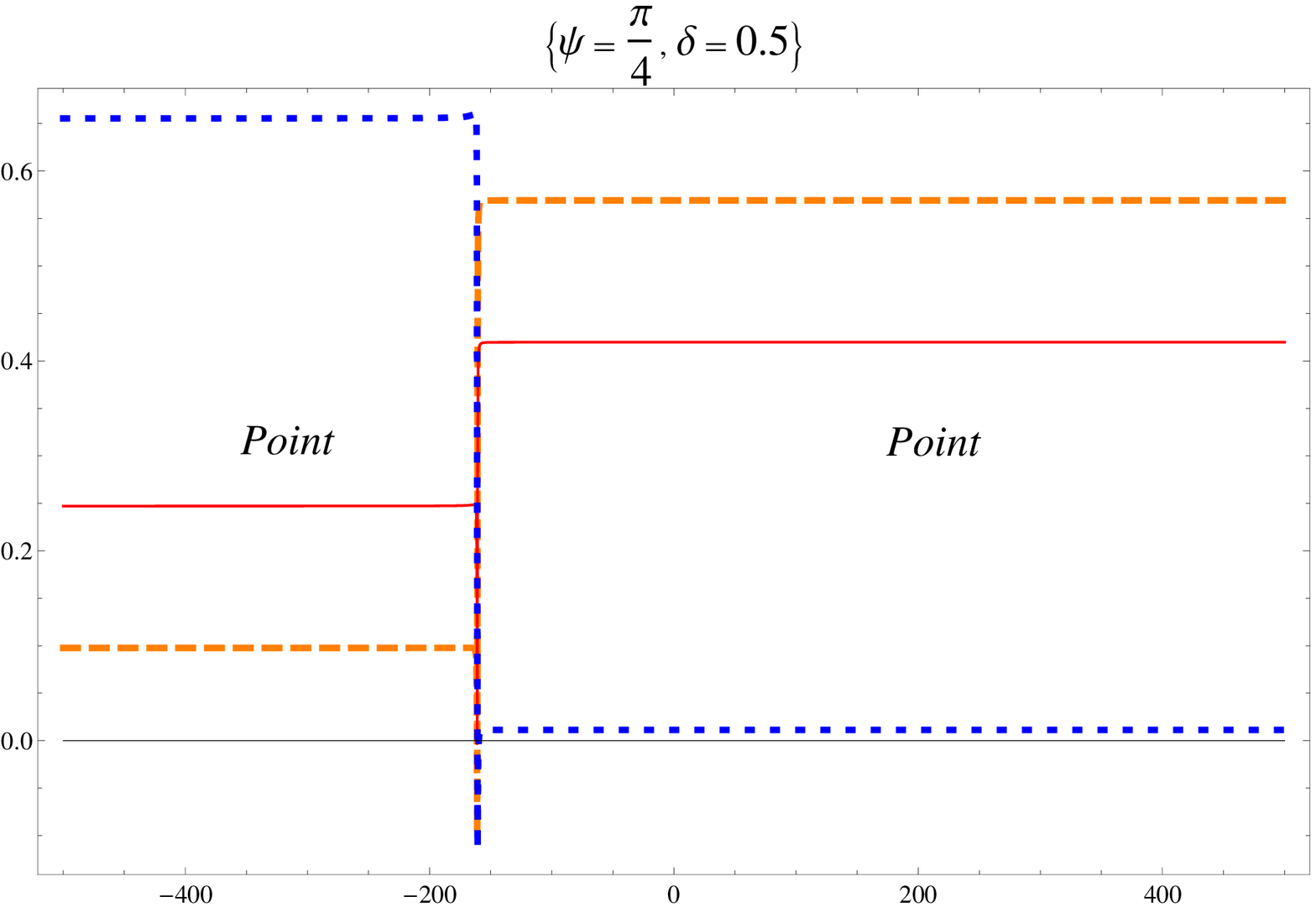}
\caption{The left figure shows the axisymmetric transition for $\delta=0.5$ and right figure shows the transition for the same value of $\delta$ and $\psi=\pi/4$ which gives the point like to point like transition.}
\label{del05}
\end{figure}

\begin{figure}[tbh!]
\includegraphics[angle=0,width=0.45\textwidth]{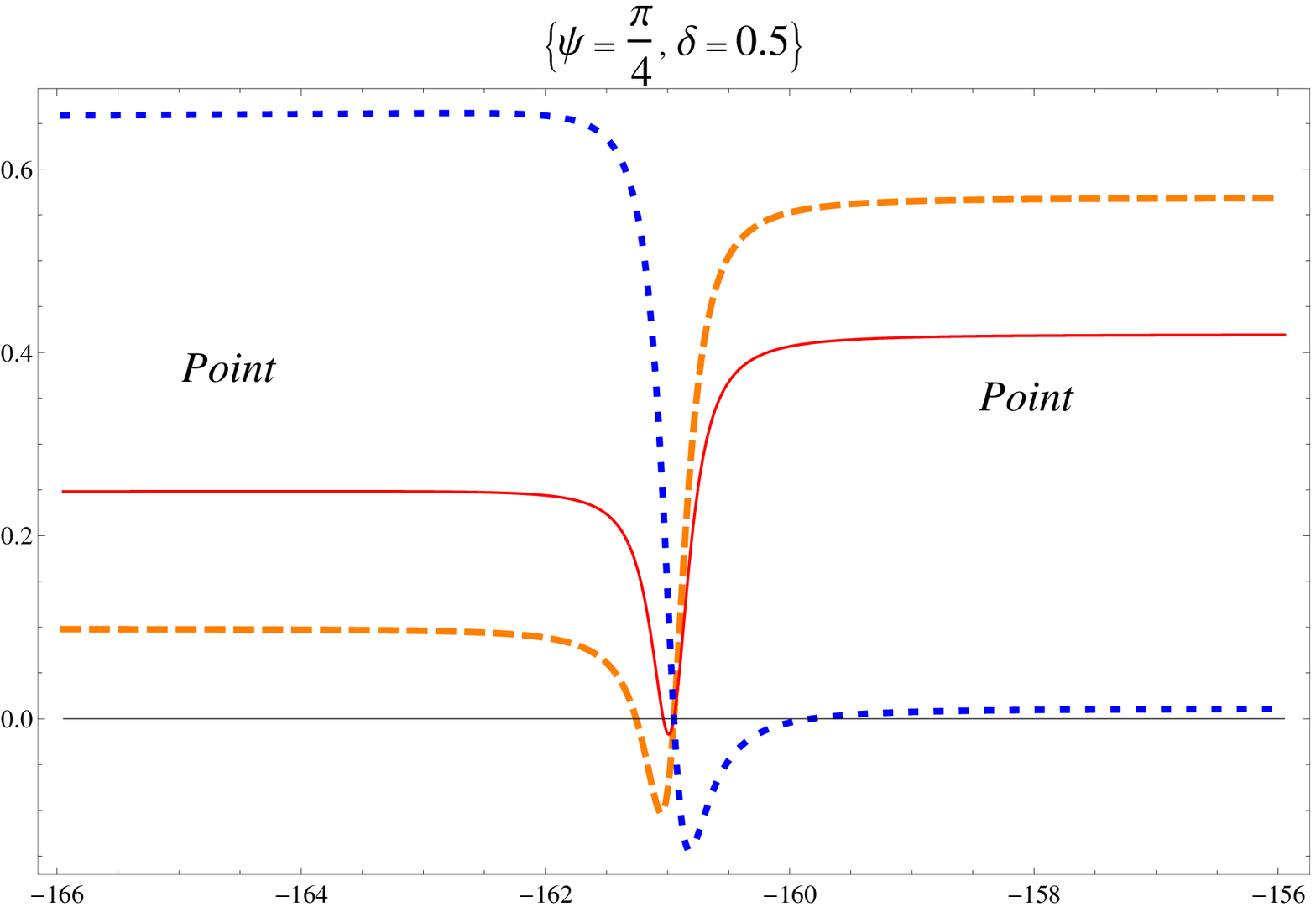}
\caption{This figure shows the transition of Kasner exponents (from point like to point like) zoomed in near the bounce for $\delta=1/\sqrt{3}$ and $\psi=2\pi/3 -0.1$. }
\label{05zoom}
\end{figure}

This range of $\delta$ is highly isotropic range in the sense that the matter dominates over the anisotropy so much that the transition is always point like to point like except for the two special cases namely $(\delta= 1/2,\, \psi=\pi/2)\,{\rm and}\,(\delta=-1/2,\, \psi=\pi/6)$, for which the transition is point like to barrel. The barrel to point like transitions again appear at a fixed value of $\psi$. This feature is shown in \fref{del05}. In \fref{05zoom} the transitions for $\delta=1/2,\,{\rm and}\, \psi = \pi/4$ has been zoomed in near the bounce. The transition involving barrel is the axisymmetric case and all the other transitions are always point like to point like transitions. For $|\delta|<1/2$ the only structure possible is a point like and therefore the transition for $|\delta|<1/2$ is always point like to point like. As one decreases $|\delta|$ further towards zero, the universe becomes more and more isotropic.

\begin{table}[tbh!]
	{\textbf{Table-II}} \\[3pt]
	\begin{tabular}{|c|c|c|}
	\hline
	&  Expanding branch & Contracting branch\\[5pt]
	\hline
$|\delta| = 1/2$ 	&		Point                                                         &    Point $\psi \neq \f{\pi}{2}$\\
				&        $\psi \neq \f{\pi}{6} \, (\delta > \,0)$          	&    Barrel $\psi = \f{\pi}{2}$ \\\cline{2-3}
	
				&         Point                                                        & Point $\psi \neq \f{\pi}{6}$ \\
				&        $\psi \neq \f{\pi}{2} \, (\delta < \,0)$          &  Barrel $\psi = \f{\pi}{6}$ \\\cline{2-3}
	\hline
$0< |\delta| < 1/2$		&	Point     ($\forall \, \psi$)                                                    &    Point ($\forall \, \psi$) \\
 	\hline
	
$-1/\sqrt{3}\, < \,  \delta \, < \, -1/2$  &   Point                                  &    Point $0 \leq \, \psi <\,\psi_c- \f{\pi}{3}$\\
				&        $0 \, \leq \, \psi <\, \psi_c$            		&    Barrel $\psi=\psi_c- \f{\pi}{3}$ \\
	                         &                                        &    Cigar $\psi_c- \f{\pi}{3} <\, \psi < \, \f{2\pi}{3}-\psi_c$ \\
	                          &                                      &    Barrel $\psi=\f{2\pi}{3}-\psi_c$ \\
	                          &                                      &    Point $\f{2\pi}{3}-\psi_0 < \, \psi < \, \psi_c$\\\cline{2-3}
	
						       &	 Barrel                                                        & Point \\
							&        $\psi=\psi_c$                             		& $\psi=\psi_c$\\\cline{2-3}
	
							&	Cigar 							& Point \\
							&	 $\psi_c < \, \psi  < \, \pi-\psi_c$ 		& $\psi_c < \, \psi  < \, \pi-\psi_c$ \\\cline{2-3}
	
							&	Barrel 							& Point \\
							&	$\psi=\pi-\psi_c$					& $\psi=\pi-\psi_c$\\\cline{2-3}
	
							&        Point  							& Point \\
							&        $\pi-\psi_c < \, \psi <\, \f{2\pi}{3}$		& $\pi-\psi_c < \, \psi < \, \f{2\pi}{3}$\\\cline{2-3}
	\hline
	
$ \delta=-1/\sqrt{3}$		&		Barrel  ($\psi =0$)                                                        &    Barrel ($\psi=0$)\\\cline{2-3}
 
	    &    Point ($0 < \psi < \f{\pi}{3}$)                                                        & Cigar ($0 < \psi < \f{\pi}{3}$) \\\cline{2-3}
	
	&	Barrel  ($\psi =\f{\pi}{3}$)							& Barrel ($\psi =\f{\pi}{3}$) \\\cline{2-3}
	
	&	Cigar ($\f{\pi}{3} <\psi < \f{2\pi}{3}$)							& Point ($\f{\pi}{3} <\psi < \f{2\pi}{3}$) \\\cline{2-3}
	\hline
	
$-1\, < \,  \delta \, < \, -1/\sqrt{3}$ 	&		Cigar                                                        &    Cigar \\
					&        $0 \, \leq \, \psi <\, \f{\pi}{3}-\psi_c$          	&    $0 \, \leq \, \psi <\, \f{\pi}{3}-\psi_c$  \\\cline{2-3}
	
					&        Barrel                                                     & Cigar \\
					&        $\psi=\f{\pi}{3}-\psi_c$                           & $\psi=\f{\pi}{3}-\psi_c$ \\\cline{2-3}
			
					&	Point							       & Cigar \\
					&	$\f{\pi}{3}-\psi_c < \, \psi < \, \psi_c $       & $\f{\pi}{3}-\psi_c < \, \psi < \, \psi_c $ \\\cline{2-3}
	
					&	Barrel 							& Cigar \\
					&	$\psi=\psi_c$						&$\psi=\psi_c$\\\cline{2-3}
	
					&        Cigar   							& Cigar $\psi_c < \, \psi < \, \f{2\pi}{3}-\psi_c$ \\
					&  $\psi_c < \, \psi < \, \f{2\pi}{3}$			& Barrel $\psi =\f{2\pi}{3}-\psi_c$\\ 
        					&									& Point $\f{2\pi}{3}-\psi_c < \, \psi < \, \f{\pi}{3}+\psi_c$ \\
					&									& Barrel $\psi=\f{\pi}{3}+\psi_c$  \\
					&									& Cigar  $\f{\pi}{3}+\psi_c < \, \psi < \, \f{2\pi}{3}$\\\cline{2-3}
	\hline
	\end{tabular}
	\caption{Transition of structures across the bounce for ($-1<\delta\leq1/2$)}
        \label{table2}
\end{table}

\begin{figure}[tbh!]
\includegraphics[width=0.6\textwidth]{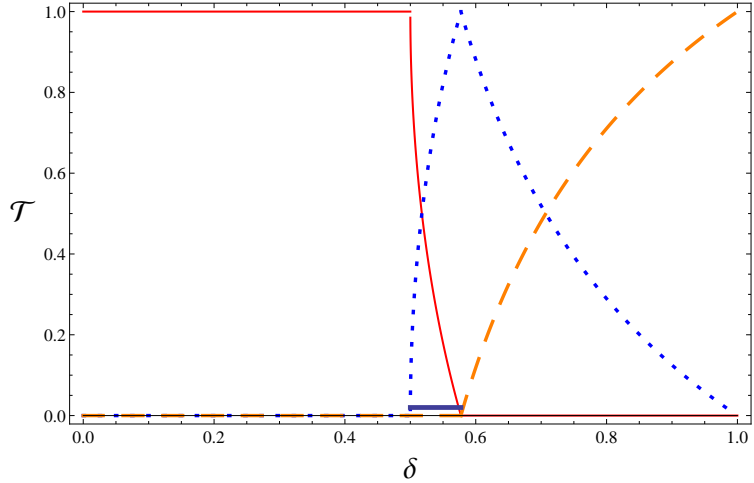}
\caption{Fraction of various transitions. The (red) solid line corresponds to point like to point like transition, (blue) dotted line represents the cigar to point like transitions and the (orange) dashed line shows the likelihood of the cigar-cigar transition. The (blue) thick line for $0.5\leq |\delta| \leq 0.57$ shows that the transitions involving barrel take place in this range, however the transition fraction $\cal T$ corresponds to a set of measure zero for such transitions.}
\label{likelihood}
\end{figure}

\begin{table}[tbh!]

	\begin{tabular}{|c|c|c|c|c|}
	\hline
	$0 < |\delta| < \f{1}{2}$ &$|\delta|=\f{1}{2}$ & $\f{1}{2} < |\delta| < \f{1}{\sqrt{3}}$ &  $|\delta|=\f{1}{\sqrt{3}}$ & $\f{1}{\sqrt{3}} < |\delta| < 1$\\[5pt]
	\hline
       P $\leftrightarrow$ P  	&	P $\leftrightarrow$ P 	&	 P $\leftrightarrow$ P 	&	P $\nleftrightarrow$ P  	&	 P $\nleftrightarrow$ P \\
      B $\nleftrightarrow$ P   	&	B $\leftrightarrow$ P 	&	 B $\leftrightarrow$ P 	&	 B $\nleftrightarrow$  P	&	B $\nleftrightarrow$  P \\           
      C $\nleftrightarrow$ P	&	C $\nleftrightarrow$ P	&	 C $\leftrightarrow$ P 	& 	C $\leftrightarrow$ P   	&		P $\leftrightarrow$ C  \\
        B  $\nleftrightarrow$ B	&	B  $\nleftrightarrow$ B 	&	 B $\nleftrightarrow$ B	& 	 B $\leftrightarrow$ B	&	B $\nleftrightarrow$ B\\
       B  $\nleftrightarrow$ C	&	B  $\nleftrightarrow$ C 	&	 B $\nleftrightarrow$ C 	& 	B $\nleftrightarrow$ C	&	 B $\leftrightarrow$ C \\
     C  $\nleftrightarrow$ C   		&	C  $\nleftrightarrow$ C	&	 C $\nleftrightarrow$ C  	&	C $\nleftrightarrow$ C	&	C $\leftrightarrow$ C\\        
    	\hline	

	\end{tabular}	
	\caption{This table shows that depending on $\delta$, some transitions are favored over others. As $|\delta|$ increases, the transitions across the bounce involve more anisotropic structures. For $|\delta|<1/2$ only transition possible is point like to point like and as $|\delta|$ increases more anisotropic transitions take place and for $|\delta|>1/\sqrt{3}$ the point like to point like transition is forbidden.}
	\label{table3}
\end{table}

\vskip0.5cm
Tables \ref{table1} and \ref{table2} summarize the geometric structures on both sides of the bounce for various values of $\delta$ in the range $-1<\delta<1$. One can see that as the value of $|\delta|$ increases, more anisotropic transitions start showing up i.e. for $|\delta|<1/2$ only point like to point like transitions are allowed. For $|\delta|>1/\sqrt{3}$ most prominent transition is cigar-cigar. In this way, there is a preference of certain transitions over the others depending on the value of $|\delta|$. Table-\ref{table3} shows all the allowed and forbidden transitions for various values of the anisotropy parameter $|\delta|$. Since, there are favored sets of transitions for different values of $\delta$, let us now define the ``transition fraction'' $\cal T$ of a transition as the ratio of the range of $\psi$ for which  a particular transition takes place to the total range of $\psi$. For example, the point like to point like transition takes place for $1/2 \leq |\delta| < 1/\sqrt{3}$ for the following values of $\psi$:

\be
\psi_{P\leftrightarrow P}= \bigg[\left(\psi_c-\f{\pi}{3}\right) + \left(\psi_c-\left(\f{2\pi}{3}-\psi_c\right)\right) + \left(\f{2\pi}{3}-\left(\pi-\psi_c\right)\right)\bigg] ~.
\ee
Therefore the transition fraction of point like to point like transition is
\be
{\cal T}_{P\leftrightarrow P}\,= \nonumber \f{4\left(\psi_c - \pi/3\right)}{2\pi/3} ~.
\ee
  For $|\delta|\leq 1/2$ the transition fraction of point like to point like transition is 1. Similarly, $\cal T$ for various transitions can be calculated:
 
 (i) for point like to point like transition 
 \be
{\cal T}_{P\leftrightarrow P} =  \begin{cases}
                    \f{6}{\pi}\left(\sin^{-1}\left(\f{1}{2|\delta|}\right)-\f{\pi}{3}\right) & ( 1/2<|\delta|<1/\sqrt{3} ) \\
                    1 & (|\delta|\leq1/2) \\
                    0 & ({\rm otherwise})
                                            \end{cases}
 \ee
 
(ii)  for point like to cigar transition
 \be
{\cal T}_{P\leftrightarrow C} =  \begin{cases}
                    \f{3}{\pi}\left(2  \sin^{-1}\left(\f{1}{2|\delta|}\right)-\f{\pi}{3}\right) & (1/\sqrt{3} <|\delta| < 1) \\
                    \f{3}{\pi}\left(\pi -2  \sin^{-1}\left(\f{1}{2|\delta|}\right)\right) & (1/2 < |\delta| < 1/\sqrt{3}) \\
                    0 & ({\rm otherwise})
                                            \end{cases}
 \ee
 
(iii)  and, for cigar-cigar transition
  \be
{\cal T}_{C\leftrightarrow C} =  \begin{cases}
                    \f{6}{\pi}\left(\f{\pi}{3} -  \sin^{-1}\left(\f{1}{2|\delta|}\right)\right) & (1/\sqrt{3} <|\delta| < 1) \\
                    0 & ({\rm otherwise})
                                            \end{cases}
 \ee

  \fref{likelihood} shows the way transition fraction of point like to point like transition depends on $|\delta|$. ${\cal T}$ for point like to point like transitions is highest for smaller values of $\delta$. As $|\delta|$ increases beyond $1/2$, the transition fraction of point like to point like transition decreases and cigar to point like transition starts increasing and attains a maximum. At this maximum, point like to point like transition has zero transition fraction and as $|\delta|$ is increased beyond this value, the cigar-cigar transitions begin to increase from zero while ${\cal T}$ for cigar to point like transitions decrease. As the anisotropy increases, cigar-cigar transitions become more prominent than the rest of the transitions. 
  \vskip0.5cm
 \noindent{\bf Remark 3:} As noted in this section, barrel structures are special in the sense that they occur at a particular value of $\psi$ for a given $\delta$ and so do the cigar-barrel or point like to barrel transitions. According to the definition of $\cal T$, the transition fraction for any transition involving barrel is essentially negligible (corresponds to a set of measure zero). Table-\ref{table3} shows that transition to (from) barrel is possible only for $1/2\leq|\delta|\leq1/\sqrt{3}$. The range of transitions involving a barrel is marked in the \fref{likelihood} by a (blue) thick line.
\vskip0.5cm
 \noindent{\bf Remark 4:} Tables \ref{table1} and \ref{table2} show an important symmetry regarding the value of $\delta$ and structure of the spatial geometry near the bounce. A positive (negative) $\delta$ in the expanding branch forms the same structure, for a given value of $\psi$, as negative (positive) $\delta$ in the contracting branch. For example, cigar structure is formed for $0\leq\psi\leq (2\pi/3-\psi_c)$ in both the cases when $1/\sqrt{3}<\delta<1$ in the expanding branch and $-1<\delta<-1/\sqrt{3}$ in the contracting branch. This behavior can be understood by writing $\delta$ in terms of the mean Hubble rate $H$ as $\delta = \pm \f{1}{H}\sqrt{\sigma^2/6}$. As the universe makes a transition from expanding to contracting branch, $H$ changes from positive to negative. This leads $\delta$ to change it sign across the bounce. For example, if one starts with a positive (negative) value of $\delta$ in the expanding branch, then, during the backward evolution across the bounce, $\delta$ becomes negative (positive) in the contracting branch.

 \subsection{Dust ($w=0$) and Radiation ($w=1/3$)}
In this subsection we  discuss the Kasner like transitions for the equation of state $w=0 \,{\rm and}\, w=1/3$. From eq.\ (\ref{w01}), it is clear that the behavior of anisotropy with respect to the matter energy density is different in the two limits i.e. $a\rightarrow 0$ and $a\rightarrow \infty$. Recall that in the classical theory, the shear scalar always dominates over the energy density (for $w<1$) and the universe behaves like a vacuum Bianchi-I near the singularity, whereas in the  asymptotic limit universe isotropizes giving the directional scale factors to be $a_i(t) \propto t^{3/2}$ for dust ($w=0$) and $a_i(t) \propto t^{1/2}$ for radiation ($w=1/3$). In LQC, as the expanding branch of the spacetime is evolved backwards in time the bounce of the mean scale factor happens well before the vacuum solutions are reached. Since there are no point like or barrel structures allowed for the dust and radiation filled universe, the transitions involve only pancake and cigar structures. Also, since the classical behavior of dust and radiation filled universe near the singularity is the same (i.e. the vacuum Kasner solution), they allow similar types of transitions across the bounce.
Table \ref{table5} shows all possible and forbidden transitions for dust and radiation filled Bianchi-I universe in LQC. It also turns out that other equations of state in the range $0\leq w <1$ follow the same selection rule given in the table. We find that in all the transitions the structure is cigar on at least one side of the bounce. There is no pancake-pancake transition present.

\begin{table}[tbh!]

     \begin{tabular}{|c|c|c|c|c|}
     \hline
		&		 P		&	B			&	Pc			&	C \\
    \hline
    P		&	$\times$		&      $\times$		&	 $\times$		&	$\times$ \\
     \hline
     B		&	$\times$		&	$\times$		& 	$\times$		& 	$\times$	    \\
     \hline
     Pc	&	$\times$		&	$\times$		&	$\times$		& $\checkmark$	\\
     \hline
     C		&	$\times$		&	$\times$		&	$\checkmark$	& $\checkmark$	\\
     \hline
     \end{tabular}
     \caption{Transitions for $0\leq w<1$. (P$\rightarrow$Point, B$\rightarrow$Barrel, Pc$\rightarrow$Pancake and C$\rightarrow$Cigar)}
     \label{table5}
\end{table}

For dust and radiation models, the pancake type structure comes out to be special (similar to barrel in the stiff matter universe) in the sense that it appears only as an axisymmetric solution occurring only for certain values of $\psi$ ($\psi=\pi/6$ when $\epsilon>0$ and $\psi=\pi/2$ when $\epsilon<0$ \cite{jacobs2}) for a given $|\epsilon|$. Moreover, pancake to pancake like transition is forbidden. This feature can be attributed to the change in the sign of the parameter $\epsilon$ across the bounce (similar to the sign change of $\delta$ discussed in remark 4). One can write $\epsilon = \pm 2H \sqrt{\f{3}{8\pi G \rho}-\f{1}{H^2}}$ which suggests that due to the change in the sign of mean Hubble rate during the transition from expanding to contracting branch of universe along the backward evolution, the parameter $\epsilon$ also changes the sign. Owing to this sign change in $\epsilon$, although $\psi$ does not change,
if one starts with an initial condition such that structure near the bounce is pancake on one side of the bounce, one does not obtain a pancake structure across. In other words, if one starts with $\epsilon>0$ and $\psi=\pi/6$ forming a pancake like structure near the bounce in the expanding branch, the parameters across the bounce will become $\epsilon<0$ and $\psi=\pi/6$ which is a condition for cigar like structure.

\begin{figure}[tbh!]
\includegraphics[angle=0,width=0.45\textwidth]{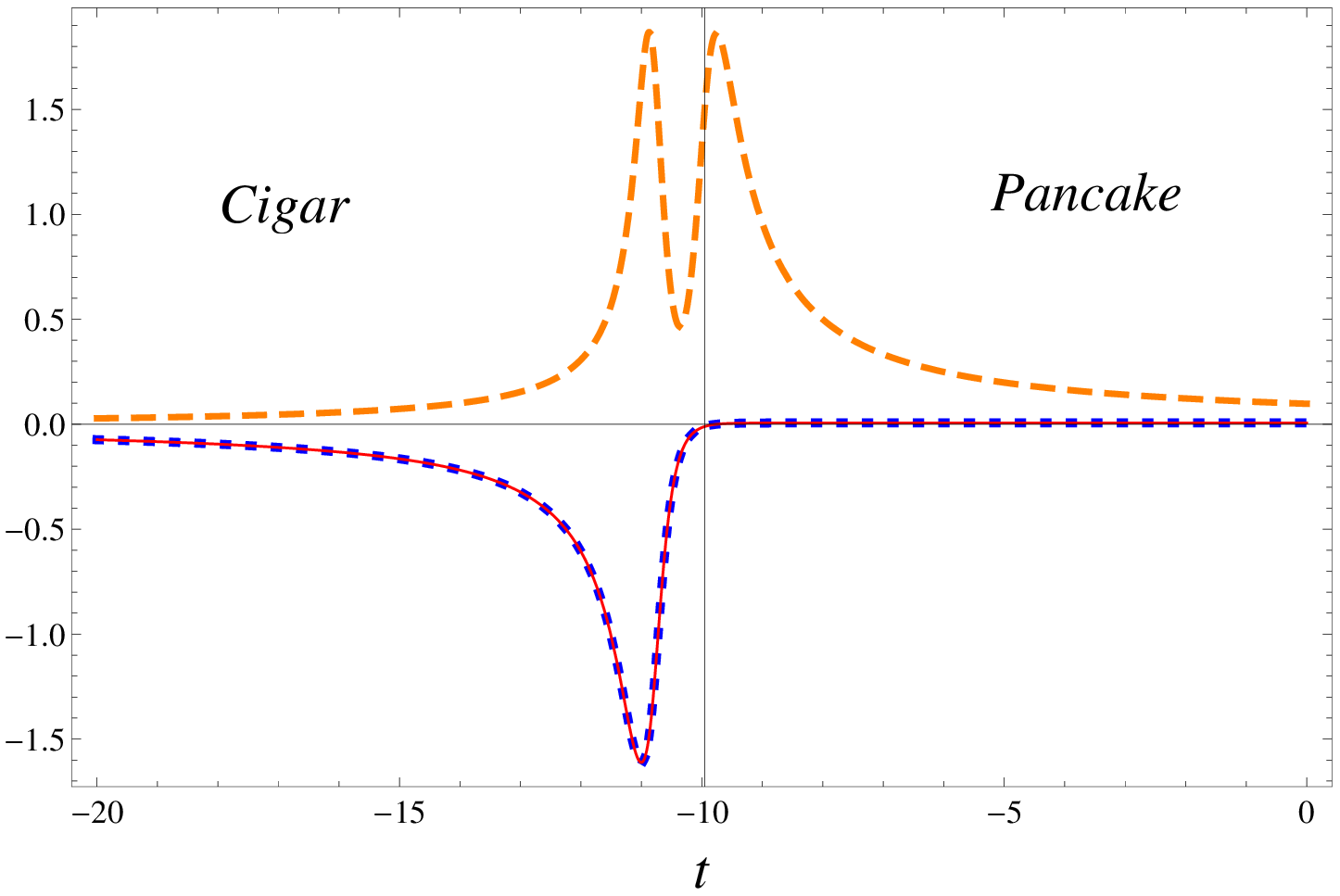}
\hskip0.5cm
\includegraphics[angle=0,width=0.45\textwidth]{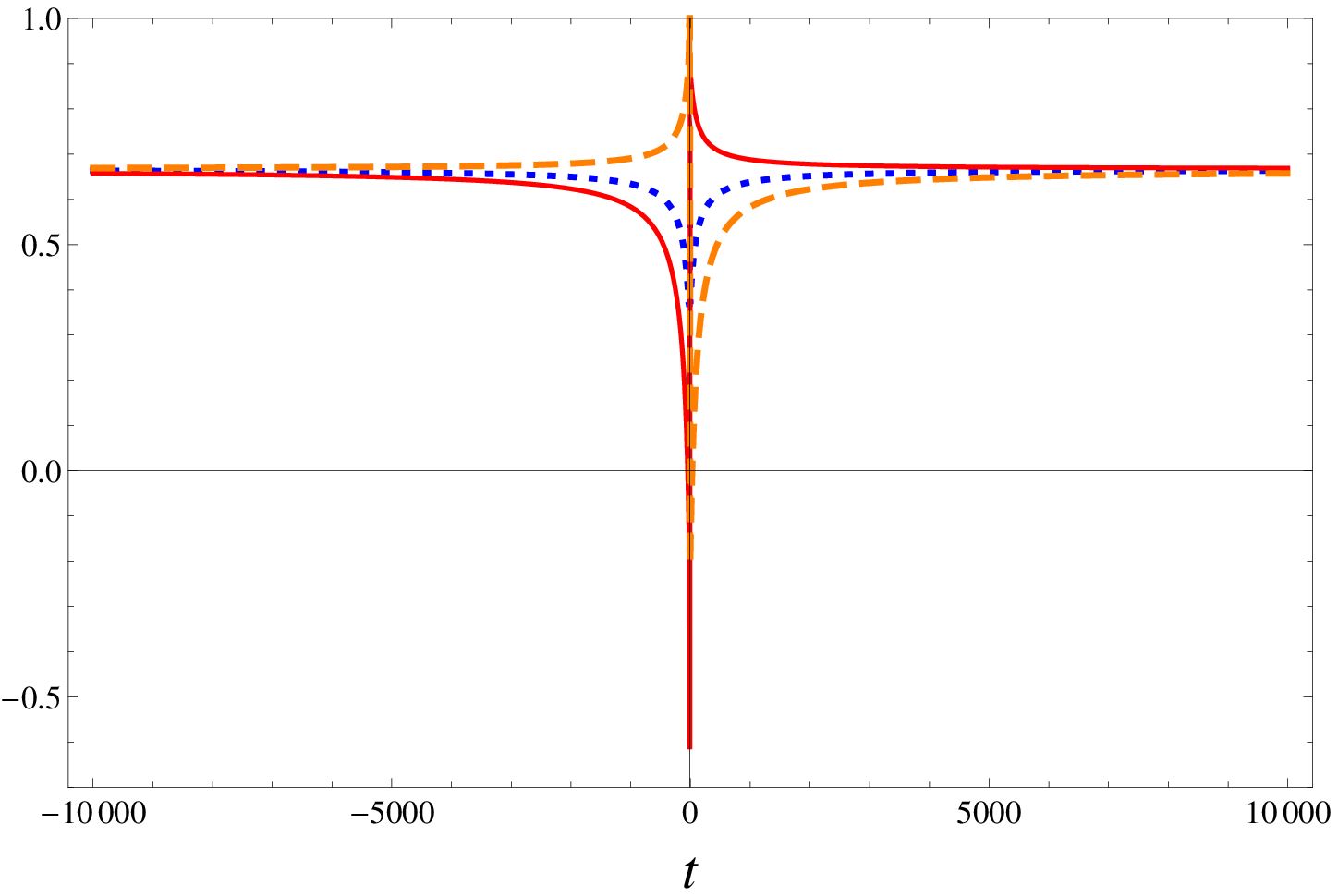}
\caption{The left plot shows the directional Hubble rates for a {\it dust} filled universe near the bounce when the spacetime undergoes a transition from a pancake type structure to a cigar. The right plot shows the evolution of the Kasner exponents far from the bounce, all the Kasner exponents tend to take the same value ($k_i=2/3$) in this limit which is an indication of the isotropization. The initial anisotropy parameters for this figure are:  $\epsilon=34 $,  $\psi=\pi/6$ (for the left plot) and $\psi=13\pi/36$ (for the right plot).}
\label{dust}
\end{figure}

\begin{figure}[tbh!]
\includegraphics[angle=0,width=0.45\textwidth]{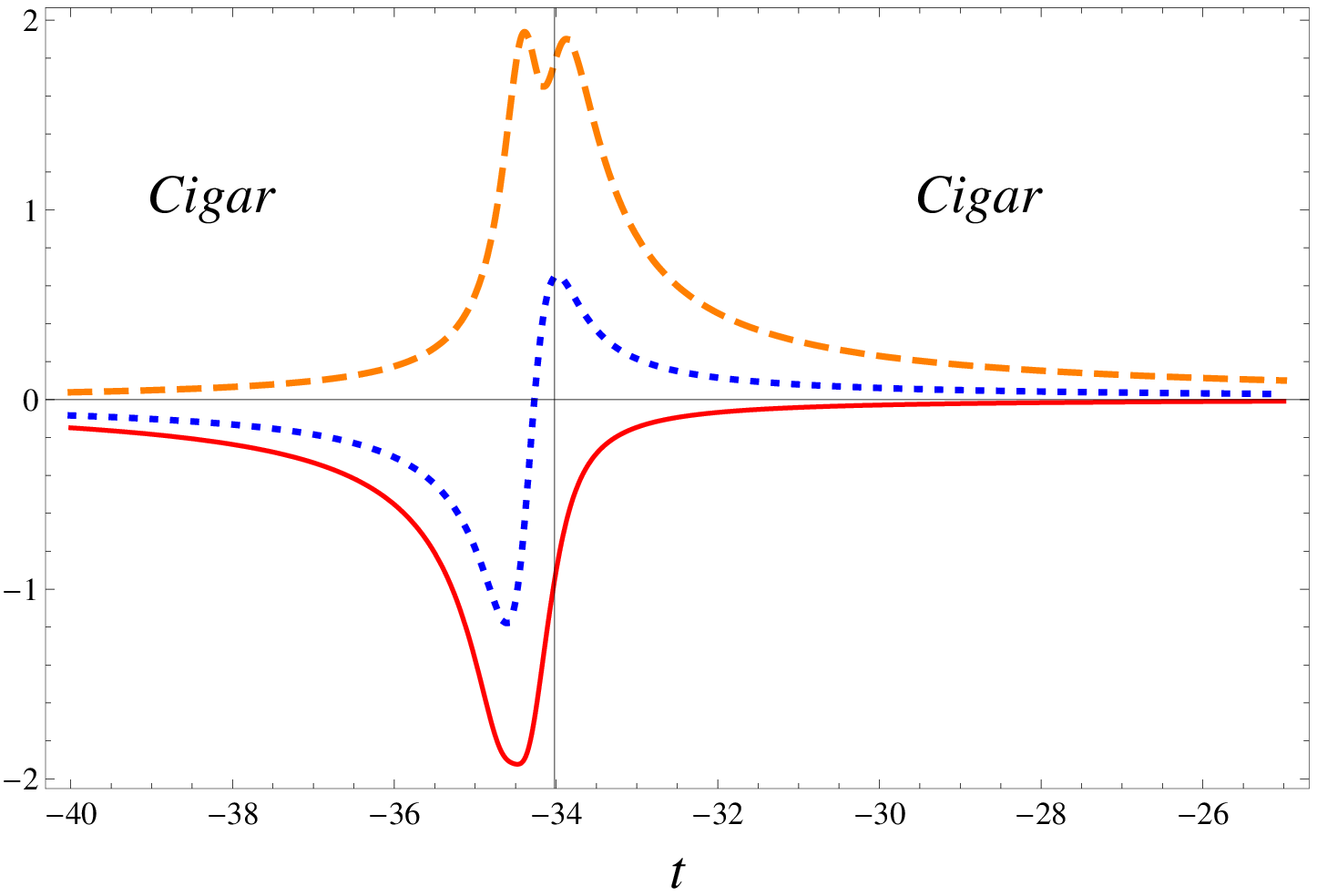}
\hskip0.5cm
\includegraphics[angle=0,width=0.45\textwidth]{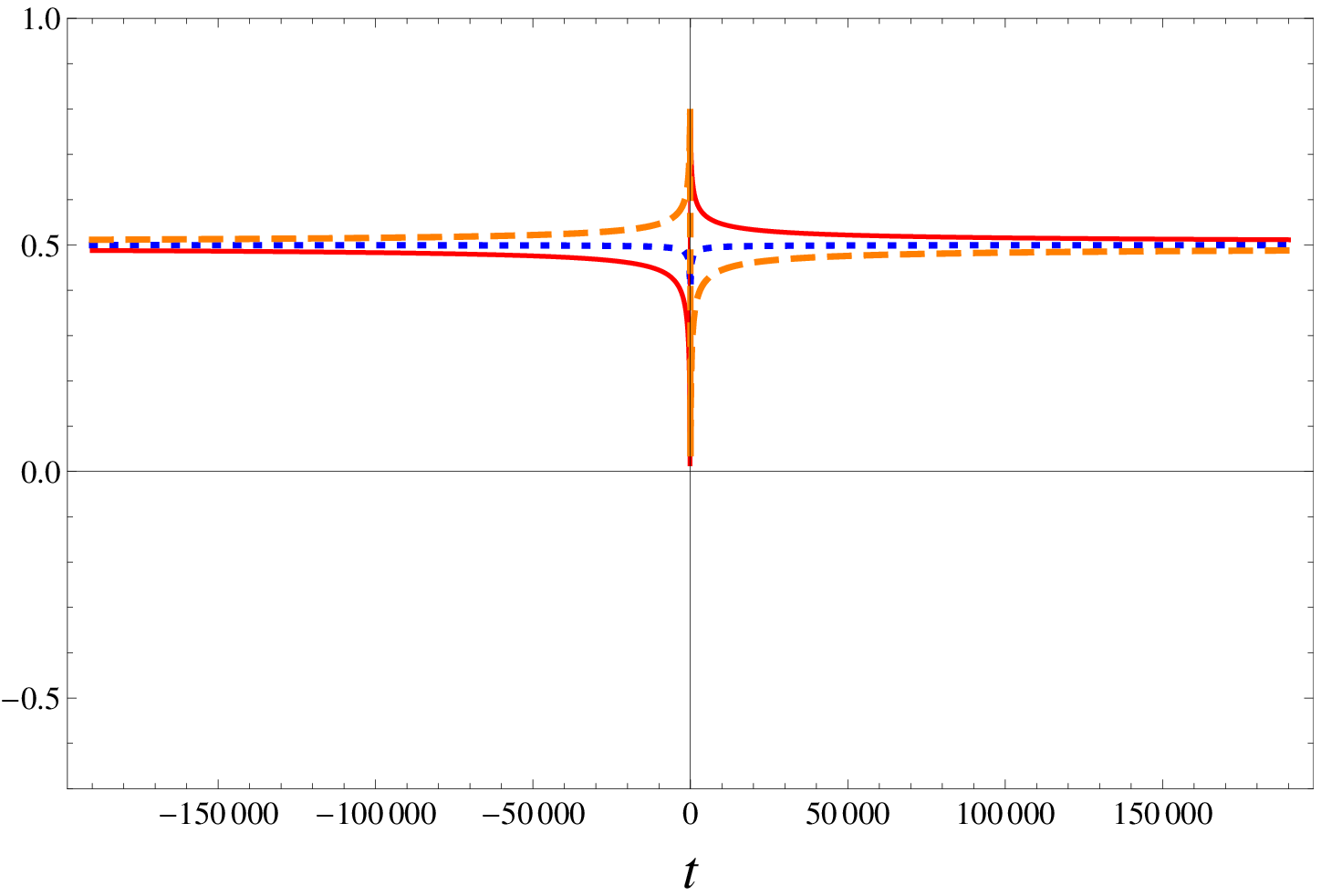}
\caption{The left plot shows the directional Hubble rates for a {\it radiation} filled universe near the bounce. The right plot shows the evolution of Kasner exponents far from the bounce when all the Kasner exponents tend to take the same value (i.e. $k_i=0.5$). This is an indication of isotropization of the spacetime at very late time. The initial conditions for this simulation are $\epsilon=8.1, \psi=\pi/18$.}
\label{rad}
\end{figure}

\fref{dust} shows the behavior of dust filled Bianchi-I spacetime across the bounce. The left plot, for which the initial parameter is taken to be $\epsilon=34\, {\rm and}\, \psi=\pi/6$, shows the directional Hubble rates. This corresponds to a pancake-cigar transition as discussed above. In the expanding branch, two of the directional Hubble rates are zero while the other is positive, this calls for a pancake type structure near the bounce.  Whereas, on the contracting side across the bounce, two of the Hubble rates are negative while the remaining one is positive leading to a cigar like structure. 
The right plot of \fref{dust} depicts the Kasner exponents with the initial anisotropy parameters being $\epsilon=34\, {\rm and}\,\psi=13\pi/36$. In this plot it is evident that in the asymptotic limit all the Kasner exponents tend to take the same value (i.e. $k_i\rightarrow 2/3$) which leads to the isotropization of the spacetime such that $a_i\propto t^{2/3}$. This is also a general asymptotic behavior of a classical dust filled Bianchi-I, as noted in the previous section.

An evolution of a radiation Bianchi-I spacetime has been shown in \fref{rad}. Left figure shows the directional Hubble rates close to the bounce and the right one portrays the late time behavior of the Kasner exponents on the two sides of the bounce. It is evident from the figure that the spatial structure near the bounce is cigar on the expanding side as well as on the contracting side which entails to a cigar-cigar transition. Similar to the dust model, the pancake-cigar transition for radiation also occurs for certain value of $\psi$ ($\psi=\pi/6$ when $\epsilon>0$ and $\psi=\pi/2$ when $\epsilon<0$). The asymptotic behavior is evident from the nature of the late time behavior of the Kasner exponents (the right plot in \fref{rad}). All the three Kasner exponents tend to take the same value i.e. $k_i\rightarrow 1/2$ which is an indication of the isotropization of the radiation Bianchi-I spacetime in the asymptotic limit. This also is a general asymptotic property of a classical radiation filled Bianchi-I universe.

\vskip0.5cm

\section{discussion}

A key question for various models in LQC is to understand various physical properties in the high curvature regime and the way they change in evolution across the bounce. Our goal in this manuscript dealt with gaining insights on this issue for Bianchi-I model using effective dynamics. Bianchi-I spacetimes are one of the simplest settings where the role of Weyl curvature on the nature of the singularities becomes transparent. The dynamics of anisotropic spacetimes has been extensively studied in the classical theory \cite{dorosh,thorne1,jacobs2,ellismac,bkl,misner2}.  In contrast to the isotopic (and homogeneous) universe, where the approach to singularity is always point like, a general approach to singularity in Bianchi spacetimes can have various geometric structure depending on the anisotropic shear and the matter content. These structures, which include barrel, cigar, pancake and a point, are labelled using the Kasner exponents in the spacetime metric. Since the formation of these structures is tied to the detailed properties of the matter and anisotropies, understanding of their formation in LQC, and their evolution across the bounce provides us with valuable hints on the change in detailed composition of the universe from one side of the bounce to the other. 

Our analysis is based on the effective spacetime description of LQC, which has been tested extensively to confirm with the underlying quantum dynamics for states which lead to a macroscopic universe at late times \cite{aps2,aps3,bp,ap,apsv,kv,szulc_b1,madridb1,dgs}. In the case of Bianchi-I model considered here, effective dynamics has provided useful insights on the boundedness of the energy density, expansion and shear scalars \cite{csv,cs09,gs1} and generic resolution of strong singularities \cite{ps11}. The modified dynamical equations obtained from the effective Hamiltonian constraint approximate the classical dynamical equations to a high order of accuracy at low curvature scales. At these scales, two observers starting from the same initial conditions, say evolving backward in the expanding branch defined by the mean scale factor,  in LQC and in GR measure the same Kasner exponents, and thus deduce the same geometric structure which is formed in the high curvature regime. The important difference being that in GR structures such as barrel, pancake, cigar and point are associated to the big bang singularity, whereas in LQC these structure are finite and non-singular. Our detailed analysis for dust, radiation and stiff matter, shows that in the evolution across the bounce, the geometric structure in the post-bounce phase can be different from the one in the pre-bounce phase. A novel result of our investigation is that there are certain selection rules between the transitions of geometric structures across the bounce. These transitions imply change in the Kasner exponents across the bounce for asymptotic observers. Many of these Kasner transitions occur in a particular range of anisotropic parameters, whereas some occur only for fine tuned values (those involving the barrel for stiff matter and the pancake for dust and radiation). Thus, given the initial conditions on the anisotropies and the matter, it is possible to identify allowed and forbidden transitions. 

We show that there is a hierarchy of different geometrical structures, and in  transitions between them. Given initial conditions on the matter and anisotropies, we find the existence of allowed and forbidden transitions in the effective spacetime description. We find a definite pattern in when a particular kind of transition is favored. For example, in the case of stiff matter for large values of anisotropic parameter $|\delta| \sim 1$, cigar to cigar transitions are found to be most favored. If the value of this parameter (which depends on initial conditions) is chosen to be smaller, the likelihood of cigar to cigar transition decreases, and becomes zero for $|\delta| \leq 1/3$. In a similar way, point like to point like transitions are forbidden for $|\delta| > 1/\sqrt{3}$, and are exclusively allowed for $|\delta| < 1/2$.  Similar behavior is deduced for all other transitions (summarized in Tables \ref{table1} and \ref{table2}), and for dust and radiation (summarized in Table \ref{table5}).  These results clearly show that transitions between the anisotropic structures are forbidden in the case of stiff matter, if anisotropic content in the universe is small. Similarly, when anisotropies are large, transition between isotropic structures do not occur. In summary, depending on the strength of anisotropies, certain transitions are favored over others, and some are forbidden.

Analysis in this article provides useful insights on the structure of the spatial geometry and their transitions across the bounce. It brings out some important differences with isotropic models in LQC, which arise due to richness of physics introduced by the interplay of the Ricci and Weyl curvature components of the spacetime curvature.  These Kasner transitions draw a similarity with the mixmaster behavior for spatially curved anisotropic models in GR but with some key differences. Unlike in GR where numerous Kasner transitions occur in the approach to general singularity, in LQC Kasner transition occurs in absence of spatial curvature and only once across the non-singular bounce. In an upcoming work, we will provide more details on the BKL like phenomena in LQC in a different model \cite{gs3}. In future, it will be interesting to analyze the Kasner transitions and selection rules using quantum constraint in LQC. Such a study will shed insights on the way quantum properties of states affect allowed and forbidden transitions across the bounce.

\acknowledgements
We would like to thank Jacobo Diaz-Polo and Jorge Pullin for useful comments and discussions. This work was supported by NSF grant PHYS1068743.

\end{document}